\documentclass{aa}  

\usepackage{graphicx}
\usepackage{txfonts}
\usepackage{lipsum}
\usepackage{subcaption}         
\usepackage{lscape}             
\usepackage{placeins}           
\usepackage{newtxtext,newtxmath}
\usepackage{tikz}
\usepackage[dvipsnames]{xcolor}

\definecolor{LinkBlue}{HTML}{4DA3FF}

\usepackage{hyperref}
\hypersetup{
  colorlinks=true,
  linkcolor=LinkBlue,   
  urlcolor=LinkBlue,    
  citecolor=LinkBlue    
}
\definecolor{lime}{HTML}{A6CE39}
\DeclareRobustCommand{\orcidicon}{%
	\begin{tikzpicture}
	\draw[lime, fill=lime] (0,0) 
	circle [radius=0.16] 
	node[white] {{\fontfamily{qag}\selectfont \tiny ID}};
	\draw[white, fill=white] (-0.0625,0.095) 
	circle [radius=0.007];
	\end{tikzpicture}
	\hspace{-2mm}
}

\foreach \x in {A, ..., Z}{%
	\expandafter\xdef\csname orcid\x\endcsname{\noexpand\href{https://orcid.org/\csname orcidauthor\x\endcsname}{\noexpand\orcidicon}}
}

\begin{document}

   \title{Cosmic web stripping and starvation of low-mass filament galaxies in TNG50}

   \subtitle{}


   \authorrunning{D. Zakharova et al.}

    \author{
    Daria Zakharova\inst{1}\orcidA{},
    Gabriella De Lucia\inst{1,2}\orcidC{},
    Benedetta Vulcani\inst{3}\orcidB{},
    Lizhi Xie\inst{1,4}\orcidL{},
    Stefania Barsanti\inst{5}\orcidS{},
    Sean McGee\inst{6}\orcidM{}, 
    }
    
    \institute{
    INAF -- Osservatorio Astronomico di Trieste, Via Tiepolo 11, I-34131 Trieste, Italy\\
    \email{dzakharovaa@gmail.com}
    \and
    IFPU -- Institute for Fundamental Physics of the Universe, Via Beirut 2, I-34151 Trieste, Italy
    \and
    INAF -- Osservatorio Astronomico di Padova, Vicolo dell'Osservatorio 5, I-35122 Padova, Italy
    \and 
    Tianjin Normal University, Binshuixidao 393, 300387 Tianjin, China
    \and
    Sydney Institute for Astronomy (SIfA), School of Physics, The University of Sydney, NSW 2006, Australia
    \and
    School of Physics and Astronomy, University of Birmingham, Birmingham, B15 2TT, UK 
    }


  \abstract
    {Galaxy properties are known to correlate with their location within the cosmic web. However, the role of filaments remains poorly understood, particularly for low-mass galaxies, which are expected to be more sensitive to environmental effects. In this work, we use the TNG50-1 simulation to investigate the properties of low-mass $8 \le \log(M_{star}/M_{{sun}}) \le 10$ galaxies in filaments and in the field, when controlling for stellar and halo mass and excluding the role of groups and clusters. We find that their integrated properties, including stellar, halo mass assembly, and quenched fractions, are similar between the two environments. However, we demonstrate that filament galaxies exhibit smaller and more asymmetric cold gas discs with respect to their field counterparts. We identify two main mechanisms driving these differences. For galaxies that entered filaments in the early Universe, during the phase of active accretion, cosmic web tidal fields modify the accretion of gas and dark matter. In some systems, accretion proceeds at rates comparable to the field but with a different geometry, leading to more tangential motions in the dark matter halo and, consequently, smaller gas discs. In others, the tidal field significantly suppresses both gas and dark matter accretion, leading to a starvation-like evolution, in which galaxies gradually exhaust their gas through star formation and can eventually quench.
    In contrast, galaxies that fall into filaments at late times can undergo cosmic web stripping, a rapid hydrodynamical removal of gas analogous to ram-pressure stripping in clusters. Our results suggest that spatially resolved gas properties are sensitive to several filament-driven environmental mechanisms.
    } 


\maketitle
\section{Introduction}
\nolinenumbers
Galaxies populate a wide range of cosmic web environments, from voids and walls to filaments, groups, and clusters, and their properties correlate strongly with the environment. Both observations and theoretical models show that galaxies inside filaments tend to be more massive, less star-forming, more elliptical, and exhibiting a mass-dependent alignment of their spin with the filament axis~\citep[e.g.][]{Tempel+2014,  Chen+2017, Laigle+2018, Kraljic+2018, Winkel+2021, Hasan+2023, Zakharova+2023, Barsanti+2025, Galarraga-Espinosa+2026}. 
From an observational perspective, the cold gas properties of filament galaxies remain debated. Some studies find that galaxies closer to filaments contain less cold gas~\citep{kuutma+2017, odekon+2018, luber+2019, Castignani+2022_gas}, while others report higher gas content in filament galaxies compared to the field~\citep{Kleiner+2017}. More recent work shows that filament galaxies, when excluding group and cluster members, exhibit no significant differences from their field counterparts~\citep{Hoosain+2024, Zakharova+2024, Vulcani_MAGNET+2026}.
\par
From a theoretical perspective, the formation of dark matter halos and of their galaxies is closely connected to the geometry and dynamics of the cosmic web~\citep{Bond+1996, Katz+2003, Keres+2005, Dekel+2009, Pichon+2011}. In this framework, galaxy growth proceeds through anisotropic accretion of matter along filaments, while the large-scale tidal field regulates and redirects these flows. This anisotropic accretion is reflected in mass-dependent correlations between galaxy angular momentum and the surrounding filamentary structure. In particular, galaxy spins are observed to align with filaments~\citep{Dubois+2014, Welker+2014, Codis_orient, Ganeshaiah_Veena+2019, Kraljic+2021, Barsanti+2022}, indicating that gas and dark matter are accreted preferentially in directions defined by the cosmic web. Halos embedded in filaments experience anisotropic tidal forces that can suppress the growth of low-mass halos while promoting the growth of more massive systems~\citep{Lemson+1999, Mo+1996, Hahn+2007, Wang+2011, Lee+2017, Borzyszkowski+2017, Musso+2018}. As a result, the cosmic web can reduce or even halt the accretion of not only dark matter but also cold gas onto low-mass galaxies \citep[e.g.][]{Aragon_Calvo+2019}. 
\textcolor{black}{Recently,  \citet{Zarattini+2025}  and \cite{Galarraga-Espinosa+2026} discussed the possible imprint of the cosmic web tidal field on galaxy properties, arguing that this physical process contributes to the reduced star-formation activity observed in filament galaxies at fixed stellar mass and local density. } 
\par
Processes operating in filaments may also affect the cold gas present in galaxies. For instance, cosmic web stripping, driven by interactions between galaxies and the intra-filament medium, can efficiently remove gas from galaxies~\citep{Benitez-Llambay+2013, Vulcani+2021, Pasha+2023, Luber+2025, Benavides+2025, Wu+2026}. At the same time, filaments are expected to contain a significant fraction of the cold gas in the Universe and may therefore act as reservoirs of accreting material, enhancing gas accretion onto galaxies~\citep{Kleiner+2017, Jego+2026}, or support star formation~\citep{Vulcani+2019}. The dominant mechanism may depend on the cosmic epoch, as filaments themselves evolve. Theoretical studies show that, depending on the adopted definition, filaments become progressively thicker, denser, and more massive as initially thin structures merge into the prominent filaments observed at late times~\citep{Cautun+2014, Cadiou+2020, Zhu+2021, galarraga-espinosa+2024, Bahe+2025}. As their dark matter and gas content increase, the conditions within filaments evolve, potentially changing the balance between processes that regulate gas inside galaxies. Also, galaxies spend different amounts of time within filaments, ranging from a few to several billion years~\citep{Zakharova_envI}, and the relative importance of different processes may depend on the time spent inside filaments. Finally, filaments are not homogeneous environments, and contain a mixture of galaxies, including those associated with groups as well as more isolated systems~\citep{Tempel+2014_perls, Sarron+2019, Castignani+2022_catalogue, Zakharova+2024, Yu+2025}. 

In a recent work \citep{Zakharova_envI}, we used predictions from the semi-analytic model GAlaxy Evolution and Assembly~\citep[GAEA;][]{De_Lucia+2024} and the hydrodynamical simulation Illustris TNG100-1~\citep{Pillepich+2018} to reconstruct the environmental histories of galaxies in filaments at z=0.  We showed that up to 80\% of filament members experienced group processing during their evolution, highlighting the importance of group-driven evolution in shaping the observed properties of filament galaxies. Even when explicitly excluding group-related processing, both models predict that filaments affect galaxies. Massive galaxies that entered filaments at early times tend to gain more stellar mass through mergers than their field counterparts, consistent with expectations from enhanced accretion on massive halos, with the effect being more pronounced for earlier infall into filaments. In contrast, low-mass galaxies show no detectable differences in stellar mass assembly or quenched fraction compared to the field in both TNG100-1 and GAEA, despite theoretical expectations of suppressed growth at the low-mass end due to the tidal field of the cosmic web.
\par
In this work, we aim to better characterise the physical mechanisms through which filaments can affect low-mass galaxies. We focus on the cold gas component, which relates to gas accretion onto galaxies and is sensitive to environmental processes such as gas removal through cosmic web stripping. At the same time, cold gas is observationally accessible, enabling a direct comparison between predictions of current work and future surveys. We restrict our analysis to low-mass galaxies, $8 \le \log(M_{star}/M_{sun}) \le 10$, as these systems are expected to be more sensitive to environmental effects due to their shallow potential wells. For our analysis, we use the high-resolution hydrodynamical simulation TNG50-1. We reconstruct the full environmental histories of z=0 galaxies and exclude all systems that experienced group processing at any point in their evolution, in order to isolate the impact of the cosmic web. 
\par
This paper is organised as follows. In Section~\ref{sec:data}, we describe the model TNG50-1 and the sample selection for the analysis. In Section~\ref{sec:results}, we compare filament and field galaxies at z=0 as a function of filament infall time, and analyse the mass–size relation to characterise the differences between the populations and identify the dominant mechanisms (gas accretion, starvation, or stripping). In Section~\ref{sec:discussion}, we investigate the physical origin and epochs of these mechanisms and present an additional observational prediction based on cold-gas asymmetry. In Section~\ref{sec:conclusion}, we summarise our findings.

\section{Data}
\label{sec:data}

\subsection{The magnetohydrodynamical simulation IllustrisTNG-50}
    TNG50-1~\citep[hereafter TNG50;][]{Illustris1, Illustris2, Illustris3, Illustris4, Illustris5}  is a magnetohydrodynamical~(MHD) simulation corresponding to a comoving volume of $35^{3}~h^{-3}$ Mpc$^{3}$, dark matter particle mass of $m_{dm} \sim 4.5 \cdot 10^{5}~\rm{M}_{sun}$. The baryonic mass resolution ($m_{b} \sim 8.5 \cdot 10^{4}~\rm{M}_{sun}$) corresponds to a typical spatial resolution of $\sim 70$–140 pc in the star-forming regions of galaxies~\citep{Pillepich+2019_sfr}, which is sufficient to robustly resolve the cold gas distribution and disc sizes analysed in this work. The model adopts a $\Lambda$CDM cosmology with parameters consistent with  \cite{Planck_2016}: $\Omega_{\Lambda} = 0.6911$,  $\Omega_{m} = 0.3089$, $\Omega_{b} = 0.0486$, $H_{0} = 67.74$ km sec$^{-1}$ Mpc$^{-1}$, $\sigma_{8} = 0.8159$, $n_s$ = 0.9667. 
    \par
   TNG50 follows the formation and evolution of galaxies, including all relevant physical processes such as radiative gas cooling and heating, star formation in the dense interstellar medium, chemical enrichment from stellar evolution, galactic winds from stellar feedback, and feedback from supermassive black holes. The evolution of dark matter, cosmic gas, stars, and supermassive black holes is followed from redshift z=127 to z=0, with model outputs stored at 100 snapshots. The model parameters are calibrated to reproduce the cosmic star formation rate density, the galaxy stellar mass function and the stellar-to-halo mass relation at z=0. In IllustrisTNG, the gas component is evolved as a hydrodynamic fluid using the moving-mesh code AREPO, spanning a wide range of temperatures and densities. Star-forming gas is modelled with a subgrid multiphase prescription and forms once the hydrogen number density exceeds $n_{\mathrm{H}} \sim 0.13~\mathrm{cm}^{-3}$. Including an explicit treatment of gas dynamics, TNG50 accounts automatically for the interaction between galaxies and the surrounding medium. In particular, it includes explicit gas stripping (such as cosmic web stripping), tidal interactions, harassment, and the effects of large-scale cosmic web tidal fields.

\subsection{Sample selection}

\textcolor{black}{We followed the environmental history reconstruction framework developed in \citet{Zakharova_envI}. 
First, we identified robust filamentary structures at each snapshot within $0<z<4$ using the DisPerSE filament finder 
\citep{Sousbie_etal+2011, Sousbie+2011}. Filament identification was performed independently at each snapshot, adopting a stellar-mass threshold such that the tracer sample contains the same number of galaxies at all epochs. This corresponds to including all galaxies with $\log(M_{\rm star}/M_{sun}) > 8$ at z=4, and to a threshold of $\log(M_{\rm star}/M_{sun}) \gtrsim 9$ at z=0. This choice is motivated by the sensitivity of DisPerSE to the number density of tracers: increasing the number of tracers generally leads to the identification of a larger number of filamentary structures and to a larger total filament length. If the tracer number density were allowed to vary significantly between z=4 and z=0, some structures could be missed at earlier snapshots and recovered only at later times. In the context of our analysis, this would make the inferred filament infall times ambiguous, since a galaxy could appear to enter a filament simply because that filament was not detected at previous snapshots. We therefore aim to obtain a homogeneous filament reconstruction across cosmic time, with comparable total filament lengths at different redshifts~\citep{Malavasi+2017, Zakharova+2023}. This can be achieved either by varying the DisPerSE persistence threshold at each snapshot, as in \citet{Zakharova+2023}, or by fixing the tracer number density. We adopted the latter approach because it is simpler to implement and provides a consistent tracer sampling across snapshots. We note that \citet{Zakharova+2023} showed that limiting the tracer sample to $\log(M_{\rm star}/M_{sun}) \geq 9$ does not affect the reconstruction of filaments at z=0, since lower-mass galaxies do not efficiently trace the cosmic web. At the same time, the lower threshold adopted at z=4 ensures a sufficiently large tracer sample for robust filament identification. }
\par
We then generated ten realisations per snapshot by randomly sampling 85\% of the galaxies, and running DisPerSE on each subsample with a 5$\sigma$ persistence level. We defined the distance to the filament as the median 3D distance to the nearest filament spine across the ten realisations. We traced the main (most massive) progenitors of all low-mass $8 \le \log(M_{star}/M_{{sun}}) \le 10$ z=0 galaxies back to z$\sim$4, retaining only progenitors with  $\log(M_{star}/M_{{sun}}) > 7$. At each snapshot, we use the progenitor positions and the reconstructed
filament spines to define infall into filaments $\tau_{\rm fils}$ as the first time a progenitor crosses a distance of 1 Mpc/h from a filament. \textcolor{black}{Although filaments span a wide range of lengths, densities, and widths, we adopt a single comoving distance threshold of 1 Mpc/h to define filament membership and infall times. This choice provides a compromise between two requirements: (i) selecting galaxies located within the inner regions of filaments, where filament-related environmental effects are expected to be strongest, and contamination from field galaxies is reduced, and (ii) retaining a sufficient number of galaxies for a statistically meaningful analysis.  We tested 0.5 Mpc/h and 2 Mpc/h in \cite{Zakharova_envI}, and found that our results are stable against these choices. } We also trace the central/satellite status of each progenitor at all epochs, and retain only galaxies that remain centrals throughout their history, thereby avoiding contamination from satellite-processed and backsplash systems (this removes approximately 65\% of the low-mass galaxies at z=0). We define satellites as galaxies residing in halos of mass around $\log(M_{halo}/M_{{sun}}) \ge 10$ where they are not the central, and we require this condition to be satisfied for at least two consecutive snapshots to ensure a robust identification of infall. If a galaxy meets this criterion at any point along its history, it is excluded from the sample. We focus exclusively on two populations:
(i) Field->Filaments (FF) galaxies~(hereafter referred to as filament galaxies), defined as systems that fall into filaments from the field and remain central throughout their evolution~(avoided group processing). This population is expected to experience only filament impact. 
(ii) Field galaxies: centrals that remain farther than
$1~\mathrm{Mpc/h}$ from any filament at z=0 and have never been inside filaments or other halos. We assume this population as our control sample. Out of the 7342 low-mass z=0 galaxies in TNG50-1, our final selections identify 431 filament galaxies and 2147 field galaxies, while the remaining systems are excluded because they experienced group processing. The relative fractions of galaxies with other environmental histories (e.g. satellites and backsplash systems) are similar to those found in TNG100 and were discussed in detail in \cite{Zakharova_envI}.
\par
For each galaxy and all its progenitors, we extracted positions and velocities, gas mass~(all gas particles associated with the subhalo), stellar mass, dark matter mass and instantaneous star-formation rate associated with each subhalo. Moreover, we also 
extracted the spatial distribution of gas and dark matter particles gravitationally bound to each galaxy. We identified cold gas as gas particles with temperatures $T < 10^{5}~\mathrm{K}$, and star-forming gas as gas particles with star-formation rate $> 0$. Using gas or dark matter particles, we computed the 3D radii enclosing 90 per cent of the cold gas mass (or dark matter) and of the star-forming gas mass, denoted as $R_{90\%, cold~gas}$~($R_{90\%, dm}$) 
and $R_{90\%, SF}$, respectively. 

\par
We divided filament galaxies into four bins according to their infall time into filaments: $\tau_{\mathrm{fils}} < 3$, $3 < \tau_{\mathrm{fils}} < 6$, $6 < \tau_{\mathrm{fils}} < 9$, and $\tau_{\mathrm{fils}} > 9$ Gyr ago. For each bin, we constructed a control sample of field galaxies matched in stellar and halo mass at z=0, in order to isolate environmental effects beyond mass-driven trends. \textcolor{black}{For each filament galaxy, we randomly selected one field galaxy with a stellar mass and halo mass within 0.1 dex of the filament galaxy at z=0~($|\Delta \log (M_{\rm star} / M_{sun})| \leq 0.1$ and 
$|\Delta (\log M_{\rm halo}/ M_{sun})| \leq 0.1$). We then assigned to each matched field galaxy the infall time of its corresponding filament galaxy, thereby constructing a control sample with the same infall-time distribution as the filament population~('mock infall times').  The adopted matching tolerance of 0.1 dex was tested using a Kolmogorov-Smirnov (KS) test, which confirms that the stellar- and halo-mass distributions of the filament and matched field samples at z=0 are statistically indistinguishable.} This procedure was repeated 100 times to account for uncertainties in the mass-matching procedure. The matching was performed independently within each $\tau_{\mathrm{fils}}$ bin. Enforcing a consistent matching across all bins would ensure identical mass distributions, but reduce the sample size. We therefore adopted independent matching to maximise statistics, and verified that this choice does not affect our results. \textcolor{black}{The number of filament galaxies used in the analysis for each infall-time bin is shown in Fig.~\ref{fig:dx_evolution}. In each realisation, the number of field galaxies is matched to the corresponding number of filament galaxies. }
\par
We confirm in Appendix A that the results of \cite{Zakharova_envI} -- that filament and field galaxies exhibit indistinguishable stellar and halo mass assembly histories -- also hold for low-mass galaxies in TNG50-1. The similarity in stellar and halo mass assembly histories allows us to compare filament and field galaxies at any epoch over 0 < z < 4, even though the matching is performed at z=0.

\section{Results}
\label{sec:results}

\begin{figure}
    \centering
    \includegraphics[width=1\linewidth]{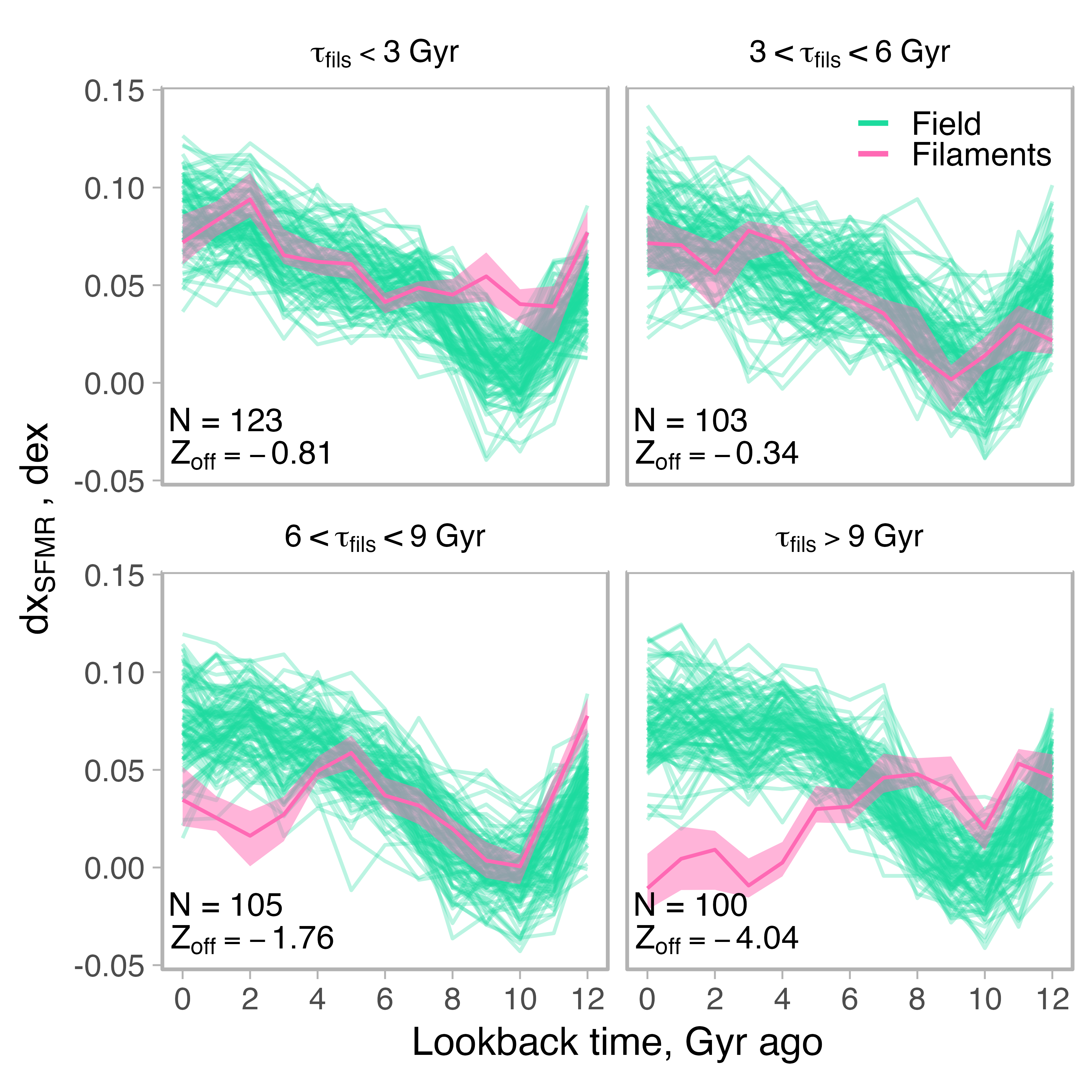}
    \caption{Deviation, $\rm dx_{SFMR}$, from the star-formation rate - stellar mass relation (estimated for each snapshot separately) as a function of cosmic time for present-day low-mass galaxies. Each panel refers to different infall times into filaments; different colours are for field and filament galaxies (for the field, we show the 100 stellar-mass–matched realisations). \textcolor{black}{The lower left corner of each panel indicates the number of filament galaxies used in the analysis after mass matching in each time bin and $Z_{\mathrm{off}}$, defined as the offset between filament and field~(control sample) galaxies in units of the field scatter.}  }
    \label{fig:dx_evolution}
\end{figure}

\textcolor{black}{Figure~\ref{fig:dx_evolution} shows the median logarithmic offset, $\mathrm{dx}_{\rm SFMR}$, from the star-forming main sequence. We define the main sequence at each snapshot considering star-forming low-mass galaxies\footnote{We note that we include the satellite population in the definition of the star-forming main sequence, which leads to a positive $\mathrm{dx}_{\rm SFMR}$ for the field population. We verified that the exact definition of SFMR does not affect the results discussed in this section. } with $sSFR > 0.15 / t_{H}$\textcolor{black}{, where $t_H =1/H(z)$ is the Hubble time at the redshift of the corresponding snapshot. At z=0, this threshold corresponds to $sSFR > 10^{-11}\,{\rm yr}^{-1}$.} Then, we compute the offset as 
$\mathrm{dx}_{\rm SFMR} = \log {\rm SFR} - \log {\rm SFR}_{\rm MS}(M_{\rm star})$. The figure shows that the median $\mathrm{dx}_{\rm SFMR}$ remains close to zero for both filament and field galaxies at all infall bins, indicating that galaxies in both samples remain consistent with the star-forming main sequence throughout their evolution. In contrast, galaxies that entered filaments at $6 < \tau_{\mathrm{fils}} < 9$ and $\tau_{\mathrm{fils}} > 9$ Gyr ago lie below the field $\rm dx_{SFMR}$. The suppression in $\rm dx_{SFMR}$ for galaxies with $6 < \tau_{\mathrm{fils}} < 9$ emerges around $\sim 4$~Gyr ago, while for $\tau_{\mathrm{fils}} > 9$ it appears earlier, $\sim 5$~Gyr ago. We quantify the present-day (lookback time $t_{\rm lb} = 0$, corresponding to z=0) in units of the field\footnote{We adopt the normalisation because the field galaxies serve as our reference sample, so that $Z_{\rm off}$ quantifies the deviation of filament galaxies from the control distribution, analogous to a z-score.} scatter as
$Z_{\rm off} = \frac{dx_{\rm fil} - dx_{\rm field}}{\sigma_{\rm field}}$,
where $dx_{\rm fil}$ and $dx_{\rm field}$ are the median deviations for filament and field galaxies, respectively, and $\sigma_{\rm field}$ is the scatter of the field population.
These values are indicated in the lower left corner of each panel.
For $\tau_{\mathrm{fils}} > 9$, we find $|Z_{\mathrm{off}}| > 4$, indicating a highly statistically significant difference between filament and field galaxies. For $6 < \tau_{\mathrm{fils}} < 9$, $|Z_{\mathrm{off}}|$ lies between $1$ and $2$; although formally less significant, a clear offset is visible. Galaxies with $3 < \tau_{\mathrm{fils}} < 6$ and $\tau_{\mathrm{fils}} < 3$ show no significant deviation from the field with $|Z_{\mathrm{off}}| < 1$.  } 
\par
Overall, TNG50-1 reveals a mild but statistically significant suppression of star formation in galaxies that have resided in filaments for more than $\sim 6$~Gyr. This suggests a subtle environmental effect, likely acting on the gas component, that does not lead to strong suppression of star formation or gas deficiency.

\subsection{Mass-size relations for galaxies in filaments at z=0}
\label{sec:mass-r90}

The sizes of gas discs are sensitive to environmental processes~\citep[e.g.][]{Conger+2025, Vulcani_MAGNET+2026}. In the case of enhanced gas accretion inside filaments, we expect larger disc sizes for the filament population. On the other hand, if the tidal field of the cosmic web disturbs or suppresses cold gas accretion, this should limit the subsequent radial growth, potentially leading to smaller gas discs at fixed stellar mass. Finally, in stripping-like mechanisms, gas removal leads to truncated gas discs and thus smaller discs.
\par
\begin{figure}[h!]
    \centering
    \includegraphics[width=1\linewidth]{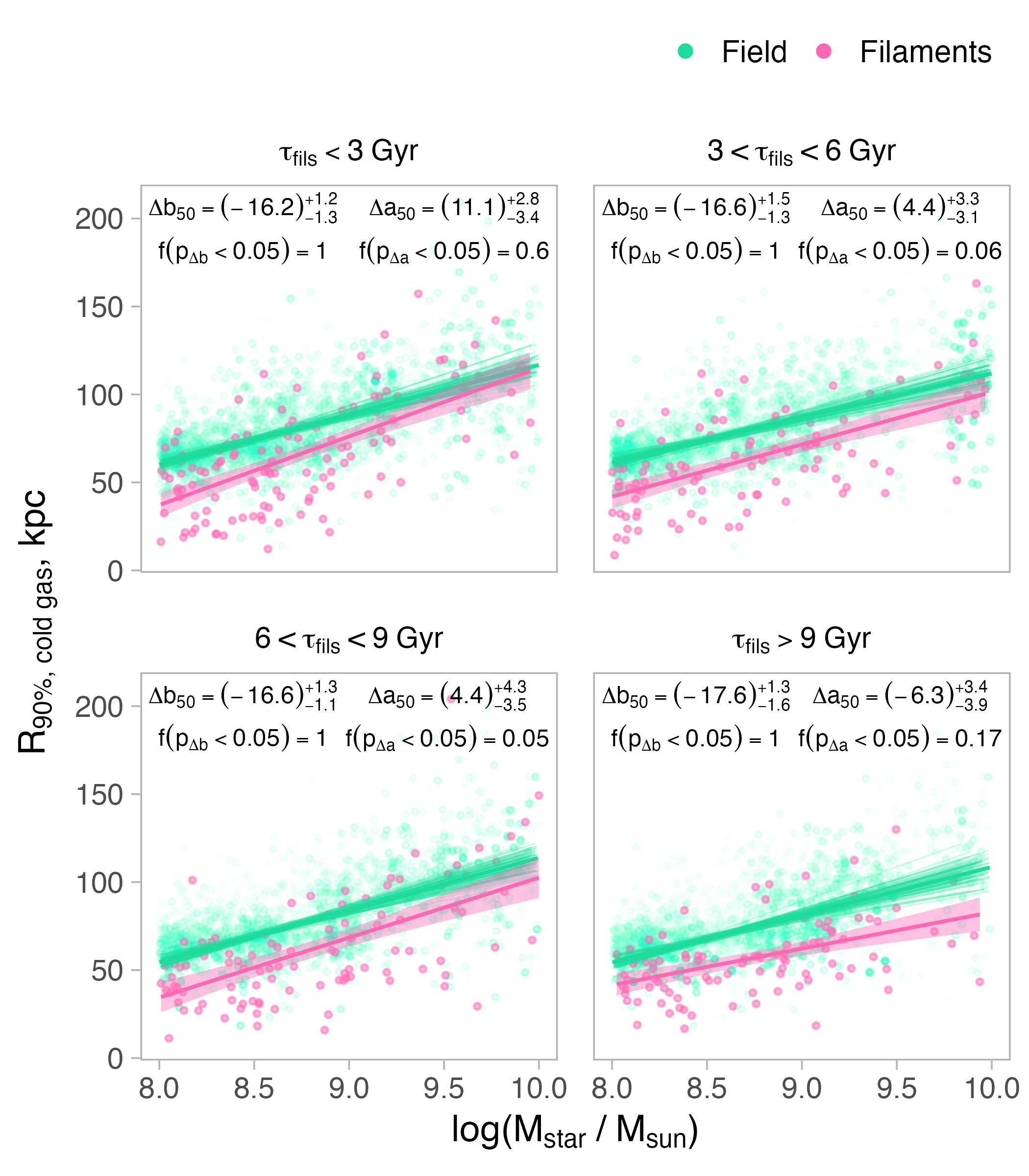}
    \caption{Mass-size $R_{90\%, cold~gas}$ relation for filament low-mass galaxies and their stellar and halo mass matched at z=0 field counterparts in bins of filament infall time $\rm \tau_{\mathrm{fils}}$. Points indicate individual galaxies. Solid lines show the best-fit linear relation for filament galaxies and for the 100 field realisations. The text in each panel reports the median differences in the linear regression coefficients (\textcolor{black}{intercept $\Delta b_{50}$, and slope ${\Delta a}_{50}$)} between filament and field samples. The second line gives the fraction of the 100 realisations for which a statistically significant difference ($p < 0.05$) between filament and field galaxies is found (see text).  }
    \label{fig:r90_cold_gas_z0}
\end{figure}

Figure~\ref{fig:r90_cold_gas_z0} shows the stellar mass--$R_{90\%, \mathrm{cold~gas}}$ relation for galaxies in filaments and in the field at z=0, split into four bins of infall time into filaments. Individual galaxies are shown as points. The cold gas extends over a wide range, from $\sim 50$ to $150$~kpc, and shows a dependence on stellar mass. Fig.~\ref{fig:r90_cold_gas_z0} shows that filament galaxies systematically lie below field galaxies in all four panels.
\par
To quantify this difference, we use an analysis of covariance (ANCOVA), which tests whether a single relation can describe both environments, or whether additional terms accounting for the environment are required. This method fits a linear relation of the form  $R_{90\%, cold~gas} = a \cdot \log (M_{star} / M_{sun}) + b$
simultaneously for both environments, treating stellar mass as a continuous covariate and environment as a categorical factor. The statistical significance of these differences is quantified through the $p$-value, which measures the probability that the filament and field galaxies can be described by the same mass-size relation. Low $p$-values indicate that separate relations are required for the two environments. For each panel of Figure~\ref{fig:r90_cold_gas_z0}, we fit the relation separately for the 100 mass-matched realisations. For each realisation, we compute the difference between the best-fitting parameters of the two relations, defined as \textcolor{black}{$\Delta b = b_{\mathrm{filaments}} - b_{\mathrm{field}}$}, and 
$\Delta a = a_{\mathrm{filaments}} - a_{\mathrm{field}}$. The corresponding p-values quantify the significance of these differences. Fig.~\ref{fig:r90_cold_gas_z0} shows the results of the linear fit as lines. The \textcolor{black}{top left corner} of each panel presents the median \textcolor{black}{over 100 realisations} difference between intercepts \textcolor{black}{$\Delta b_{50}$} with 1$\sigma$ confidence intervals and the fraction of the 100 field realisations for which the difference with respect to the filament fit is statistically significant \textcolor{black}{$f(p_{\Delta b} < 0.05)$}. The top right corner represents the same, but for median slope \textcolor{black}{${\Delta a}_{50}$ and its significance $f(p_{\Delta a} < 0.05)$.}


\textcolor{black}{The ANCOVA confirms that filament galaxies have smaller cold gas discs for $\Delta b_{50}$ from  -16 to - 18 kpc}, and this offset in the mass–size relation is always statistically significant, with $f(p_{\Delta b} < 0.05) = 1$ in all panels. Also, the panel with $\tau_{\mathrm{fils}} < 3$~Gyr shows a large fraction of realisations that have a statistically significant difference  $f(p_{\Delta a} < 0.05) = 0.6$  in the slope of the mass-size relation. The size difference between filament and field galaxies becomes less significant for $\log_{10}(M_{star}/M_{sun}) > 9.5$. This indicates that the mechanism driving the detected differences for galaxies that only recently came close to a filament is stellar-mass dependent. In contrast, for earlier infall times, the fraction of realisations with statistically significant differences in slope is low~\textcolor{black}{($f(p_{\Delta a} < 0.05) < 0.5$)}, suggesting that the processes responsible for the size differences are independent of stellar mass.
\par
In Appendix~\ref{app:mass-size50}, we show that $R_{50\%, \mathrm{cold~gas}}$ follows the same trends. This suggests that the impact of the filaments is not limited to the truncation of the outer gas, but also involves changes in the internal gas distribution, consistent with either suppressed disc growth or redistribution of gas within the disc.
\begin{figure}
    \centering
    \includegraphics[width=1\linewidth]{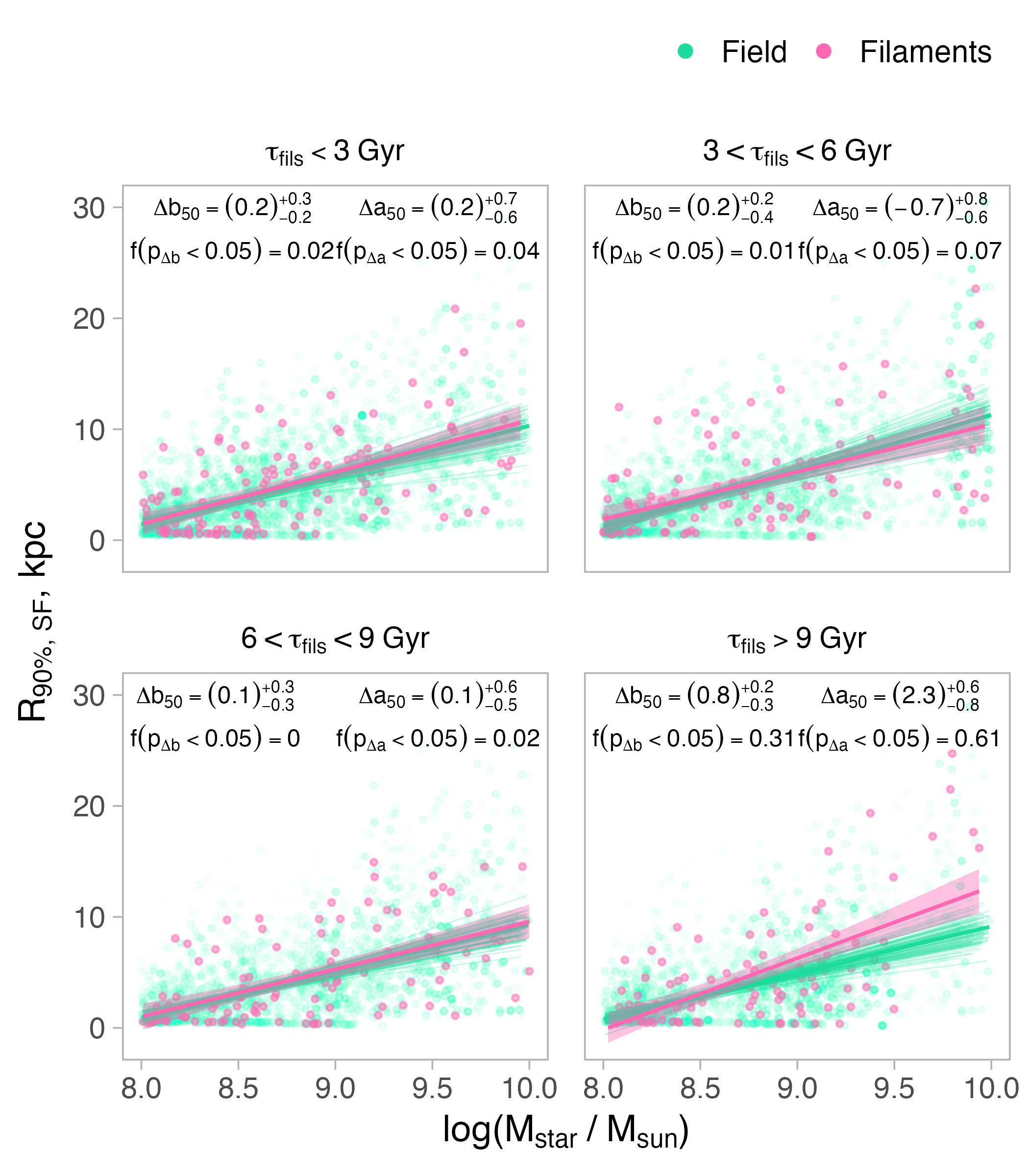}
    \caption{Same as Fig.~\ref{fig:r90_cold_gas_z0} but for 90\% of star-forming gas. }
    \label{fig:r90_sfr_gas_z0}
\end{figure}


We repeat the same analysis for the size of star-forming regions \textcolor{black}{$R_{90\%, \rm SF}$} in Figure~\ref{fig:r90_sfr_gas_z0}. We find that the star-forming gas is located in the very centre of galaxies, rarely exceeding the inner 20 kpc, both for galaxies in the field and filaments. There is no significant difference between filament and field galaxies, regardless of the time of infall into filaments. This is consistent with the absence of differences in integrated properties, indicating that star formation proceeds similarly in filament and field galaxies. Therefore, the impact of filaments is confined to regions beyond the star-forming core, primarily affecting the outer gas reservoir rather than the central star-forming regions. This finding is consistent with recent observational studies that found no significant difference in the star formation concentration index in filaments and voids~\citep{Barsanti+2026}.
\par
Taken together, we find no evidence for enhanced gas accretion in filaments, which would produce larger and more extended discs. Rather, filament galaxies exhibit systematically smaller gas discs across all infall-time bins, indicating that their radial growth is suppressed or truncated. This behaviour is consistent either with reduced gas accretion, for example, due to tidal interactions with the cosmic web, or with direct truncation of the gas disc via cosmic web stripping. We explore these possibilities in more detail below.

\subsection{Growth of gas disc size inside filaments}
\label{subsec:growth_of_disc}

To establish which processes lead to smaller gas disc sizes of filament galaxies, we examine the evolution of $R_{90\%,\mathrm{cold~gas}}$ for filament galaxies before and after infall into filaments. This allows us to determine whether galaxies initially grow their gas discs and subsequently experience rapid truncation, consistent with stripping-like processes (cosmic web stripping), or whether their disc growth is gradually suppressed due to reduced cold gas accretion driven by tidal fields of the cosmic web. \textcolor{black}{We compare the growth rate of the cold gas disc size, $R_{90\%, cold~gas}$,  
for galaxies residing in filaments before and after their infall, tracing the main progenitor, and contrasting these trends with their field counterparts to which we assign  "mock infall times" (see below).}
\begin{figure*}
    \centering
    \includegraphics[width=1\linewidth]{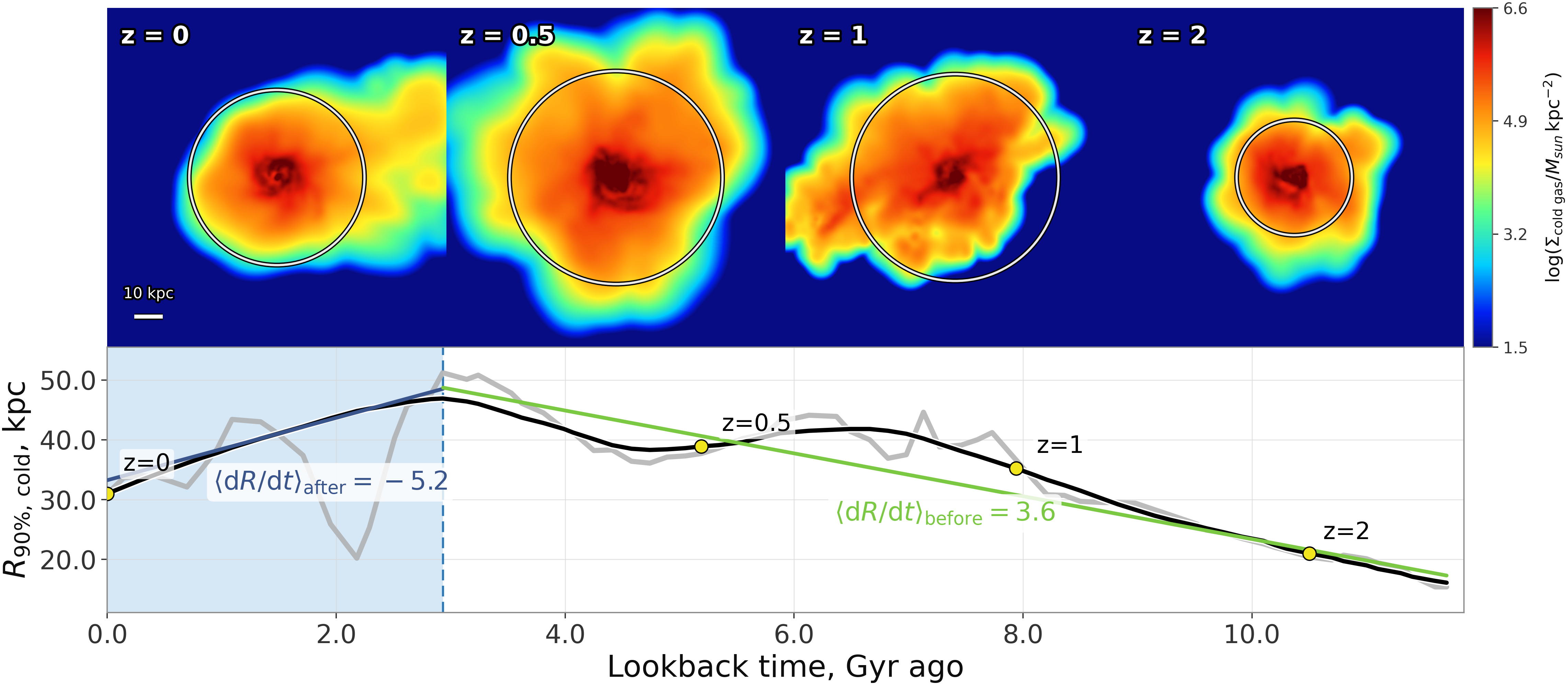}
    \caption{
    \textcolor{black}{Example of the growth rate of the cold-gas disc size $R_{90\%, cold~gas}$ for a single filament galaxy as a function of cosmic time. The top panel shows face-on maps of the cold-gas surface density for the example galaxy at z=0, 0.5, 1, and 2, with $R_{90\%, cold~gas}$ shown as white circles. The bottom panel shows the evolution of $R_{90\%, cold~gas}$ at each snapshot for an example galaxy. The thick grey curve shows $R_{90\%, cold~gas}$ estimated at each snapshot, while the black line shows its locally smoothed trend.  The evolution is analysed separately before~(white region) and after~(blue region) the galaxy’s infall into the filament by computing the mean $\mathrm{d}R/\mathrm{d}t$~(kpc/Gyr$^{-1}$) in each region. For this galaxy, the before infall phase exhibits rapid disc growth, while after infall into filaments the galaxy undergoes disc truncation. }}

    \label{fig:size_evolution_example}
\end{figure*}


\begin{figure*}
    \centering
    \includegraphics[width=1\linewidth]{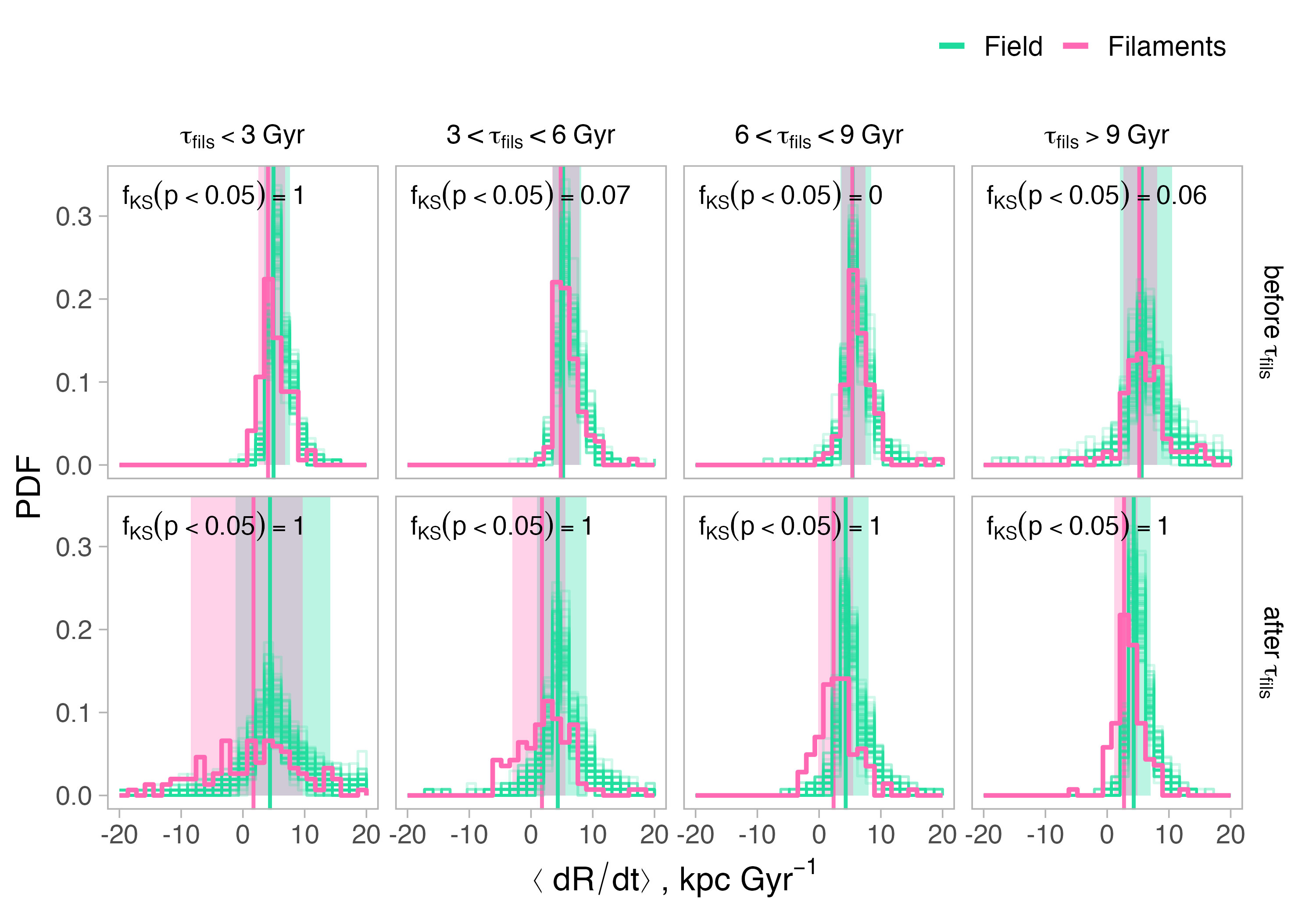}
    \caption{ 
    Growth rate of the cold gas disc radius $R_{90\%, \mathrm{cold~gas}}$ for filament galaxies compared to field galaxies, shown separately before infall (top row) and after infall (bottom row) into filaments. For field galaxies, a mock infall time is assigned, allowing the same before and after segmentation and a consistent estimate of the growth rates. All 100 stellar- and halo-mass–matched field realisations are shown in every panel. The vertical lines indicate the median growth rate and the corresponding $1\sigma$ spread. \textcolor{black}{The top left corner of each panel reports $f_{\rm KS}(p<0.05)$, the fraction of the 100 field realisations for which the KS test rejects the null hypothesis that the filament and field distributions are the same distribution.}}

    \label{fig:size_evolution_anal}
\end{figure*}

\textcolor{black}{To estimate the evolution of $R_{90\%, cold~gas}$ for a single galaxy we used the following procedure. For each galaxy, we measured $R_{90\%, cold~gas}$ at each snapshot for the main progenitors. Since $R_{90\%, cold~gas}$ can be noisy due to snapshot-to-snapshot fluctuations, we applied a LOESS regression~(locally estimated scatterplot smoothing) to obtain a smoothed representation of the radius evolution. Then, we estimated the growth rate  of gas disc $\frac{\mathrm{d}R}{\mathrm{d}t} $ as a central finite-difference derivative of the LOESS:
\begin{equation}
\left(\frac{{\rm d}R}{{\rm d}t}\right)_j
= -
\frac{
R_{\rm smooth}(t_{{\rm lb},j+1}) -
R_{\rm smooth}(t_{{\rm lb},j-1})
}{
t_{{\rm lb},j+1} -
t_{{\rm lb},j-1}
}.
\end{equation}
Afterwards, we split galaxies based on their $\rm \tau_{\mathrm{fils}}$ into two intervals before and after infall \textcolor{black}{and average the local derivatives within each
interval separately. The resulting quantities,} $\left\langle \frac{{\rm d}R}{{\rm d}t} \right\rangle_{\rm before and after}$, \textcolor{black}{measure the average radius evolution
during the corresponding phases, with positive values indicating 
radial growth and negative values indicating truncation.} Figure~\ref{fig:size_evolution_example} provides an illustrative example of this procedure for one filament galaxy. The top panels show face-on maps of the cold-gas surface density at different redshifts, with the corresponding $R_{90\%, cold~gas}$ shown as white circles. The bottom panel shows the time evolution of $R_{90\%, cold~gas}$: the grey curve corresponds to the raw snapshot values, while the black curve shows the LOESS-smoothed trend. The shaded region demonstrates the time after filament infall. The coloured segments indicate the mean growth rates measured before and after infall $\left\langle \frac{{\rm d}R}{{\rm d}t} \right\rangle_{\rm before and after}$. We note that we are interested only in the net evolution of $R_{90\%, cold~gas}$ before and after infall, rather than in identifying individual episodes of growth or truncation.  }
\par
We applied the same procedure to filament and field galaxies by assigning to the latter a mock infall time to match the distribution of filament infall times. This allows us to have before- and after-infall phases also for field galaxies, and to compute their growth rates using an identical methodology to that adopted for filament galaxies. We recall that field and filament galaxies are constructed to have the same stellar and halo mass distribution, ensuring a fair comparison between the two populations within each $\rm \tau_{\mathrm{fils}}$ bin. 
\par
The outcome of this procedure is shown in Fig.~\ref{fig:size_evolution_anal}. The first row presents the distribution of mean \textcolor{black}{$\left\langle \frac{{\rm d}R}{{\rm d}t} \right\rangle$} before infall for filament and field galaxies. \textcolor{black}{We perform a two-sample KS test between the filament sample and each of the 100 mass-matched field realisations. We report $f(p<0.05)$, the fraction of realisations for which the null hypothesis that the filament and field growth-rate distributions are drawn from the same parent distribution is rejected. This fraction therefore measures how robustly the filament growth rates differ from those of the matched field sample.} Almost all galaxies before infall exhibit positive \textcolor{black}{$\left\langle \frac{{\rm d}R}{{\rm d}t} \right\rangle$}, indicating that their gas discs grow in size before entering filaments, with a median growth rate of $\sim 4$--$6$~kpc~Gyr$^{-1}$. Filament galaxies and their field counterparts exhibit similar distributions, and the KS test does not reveal any significant difference for the disc growth rate before infall: less than $30\%$ of the realisations have $p$-values below $0.05$, except for the $\tau_{\mathrm{fils}} < 3~\mathrm{Gyr}$ bin.  The difference between $\tau_{\mathrm{fils}} < 3~\mathrm{Gyr}$ and the field galaxies may reflect environmental effects prior to filament infall. One possible explanation is that recently accreted filament galaxies were influenced by other cosmic web environments, such as walls or lower-density filamentary structures, whose tidal fields could have affected their properties before entering filaments.
\par
The bottom panels of Fig.~\ref{fig:size_evolution_anal} show the evolution of disc sizes after infall into filaments, compared to the field. The KS test reveals that galaxies inside filaments are statistically different from their field counterparts independently of the accretion time ($f(p < 0.05) = 1$). Filament galaxies exhibit a shift of \textcolor{black}{$\left\langle \frac{{\rm d}R}{{\rm d}t} \right\rangle$} towards lower and negative values, relative to the field, with a dependence on the $\tau_{\mathrm{fils}}$ bin. For filament galaxies with $\tau_{\mathrm{fils}} > 9$ the median value is $2.7 \pm 1.9~\mathrm{kpc~Gyr^{-1}}$ against $4.3\pm2.2~\mathrm{kpc~Gyr^{-1}}$ for the field sample. However, the fraction of galaxies with negative \textcolor{black}{$\left\langle \frac{{\rm d}R}{{\rm d}t} \right\rangle$} remains small ( $9 \pm 3 \%$ for filament galaxies and $0^{+0.5}_{-0}\%$ for the field), indicating that filament galaxies primarily experience a slowdown in size growth rather than disc truncation. We interpret this as evidence of a starvation-like process affecting early infallers into filaments.
\par
For galaxies that become part of a filament at later times, the fraction of galaxies with 
\textcolor{black}{$\left\langle \frac{{\rm d}R}{{\rm d}t} \right\rangle$ < 0} after infall increases from 
$9 \pm 3\%$ to $20 \pm 3\%$, $33 \pm 4\%$, and $40 \pm 4\%$ 
from the right to the left, bottom panels of Fig.~\ref{fig:size_evolution_anal}. 
The corresponding fractions for mass-matched field galaxies are 
$1.9 \pm 1.5\%$, $5 \pm 2\%$, and $16 \pm 3\%$, respectively. This excess of filament galaxies with negative \textcolor{black}{$\left\langle \frac{{\rm d}R}{{\rm d}t} \right\rangle$} indicates the presence of gas disc truncation inside filaments, becoming more pronounced for more recent infallers. This indicates a transition in the dominant physical mechanism with infall time, from suppressed growth in early infallers to increasingly important truncation.
\par
To summarise, we find that the smaller gas disc sizes of filament galaxies at z=0 arise after infall and depend on the time spent within filaments. Galaxies that entered filaments at early times ($\tau_{\mathrm{fils}} > 9~\mathrm{Gyr}$) exhibit a suppression of disc growth that is consistent with reduced gas accretion. An increasing fraction of more recent infallers suffer from a truncation of the gas disc.

\subsection{Cold gas content within filament galaxies}
\label{subsec:mass_classification}

In this section, we examine the evolution of the cold gas reservoir. 
In particular, we measure the cold gas mass $M_{90\%, \mathrm{cold~gas}}$ enclosed within $R_{90\%,\mathrm{cold~gas}}$ to complement the size evolution analysis discussed in the previous section. We classify filament galaxies according to the evolution of their cold gas mass after infall, in order to determine whether they continue to grow their cold gas reservoir or instead exhibit a decrease in cold gas mass over time. The rate of gas loss provides a key diagnostic to distinguish between a gradual decline of the cold gas reservoir expected in starvation-like scenarios, and a more rapid depletion that corresponds to a stripping-like process.

\begin{figure*}
    \centering
    \includegraphics[width=1\linewidth]{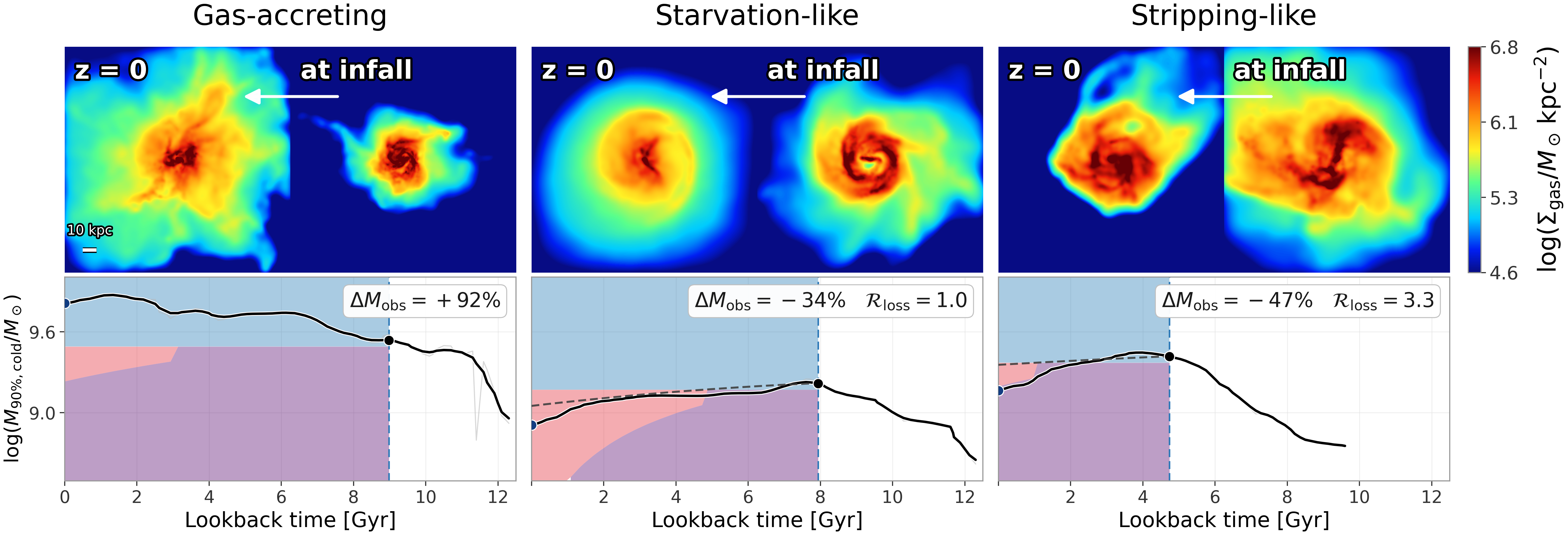}
    \caption{\textcolor{black}{ Example illustrating the method used to classify the evolution of the cold gas mass $M_{90\%,{\rm cold~gas}}$ after infall into filaments.  The classification is based on the comparison between the net change in the cold gas mass at the time of infall and at z=0. We consider three possible phases shown from left to right: gas-accreting, starvation-like, and stripping-like. Gas-accreting corresponds to cases where the cold gas mass grows or does not decrease by more than 10\% after infall (blue region). A starvation-like phase occurs when the galaxy does not accrete cold gas and gradually consumes its gas through star formation (red region). A stripping-like phase corresponds to cases where the cold gas mass decreases more rapidly than expected from star formation alone, indicating direct gas removal (purple region). The upper panels show the cold-gas surface density maps at infall and at z=0, while the lower panels show the corresponding evolution of $M_{90\%,{\rm cold~gas}}$. } }

    \label{fig:starvation_vs_stripping_example}
\end{figure*}

\par
We classify galaxies based on the evolution of their cold gas reservoir after infall into filaments as illustrated in Fig.~\ref{fig:starvation_vs_stripping_example}. For each galaxy, we measure the mass of cold gas $M_{90\%,cold~gas}$ tracing its main progenitor up to $z \simeq 4$~(grey line) and smooth over time using a LOESS regression~(black line).  For each galaxy, we compute the change in cold gas mass after infall as \textcolor{black}{$\Delta M_{\rm obs} = M_{\rm z=0} - M_{\tau_{\rm fils}}$}. Galaxies whose cold gas reservoir remains approximately unchanged or increases after infall $\Delta M_{\rm obs}/M_{\tau_{\rm fils}} \geq -0.1$, are classified as gas-accreting~(blue zone in Fig.~\ref{fig:starvation_vs_stripping_example}). If the cold gas reservoir decreases by more than 10\%, we compare the observed gas loss with the depletion expected from star formation. The reference depletion $\Delta M_{\rm ref}$ is estimated by integrating the median star formation rate of field galaxies~(stellar and halo mass matched samples) over the same time interval after infall. We then compute the ratio \textcolor{black}{$\mathcal{R}_{\rm loss} = |\Delta M_{\rm obs}| / |\Delta M_{\rm ref}|$}. If \textcolor{black}{$\mathcal{R}_{\rm loss}\lesssim 3$}, the gas decline is consistent with consumption by star formation alone and the galaxy is classified as starvation-like~(red zone in Fig.~\ref{fig:starvation_vs_stripping_example}). The threshold \textcolor{black}{$\mathcal{R}_{\rm loss} = 3$} is motivated by previous studies showing that stellar feedback in low-mass galaxies can drive gas outflows with mass loading factors of $\sim 2$–$3$~\citep[e.g.][]{Muratov+2015}. Therefore, values of \textcolor{black}{$\mathcal{R}_{\rm loss} > 3$} indicate gas loss significantly faster than can be explained by star formation alone, pointing to additional environmental gas removal, and the galaxy is classified as stripping-like~(purple zone in Fig.~\ref{fig:starvation_vs_stripping_example}). \textcolor{black}{Examples of galaxies classified using this method are shown in Appendix~\ref{app:galaxies}.}

\begin{figure}
    \centering
    \includegraphics[width=1\linewidth]{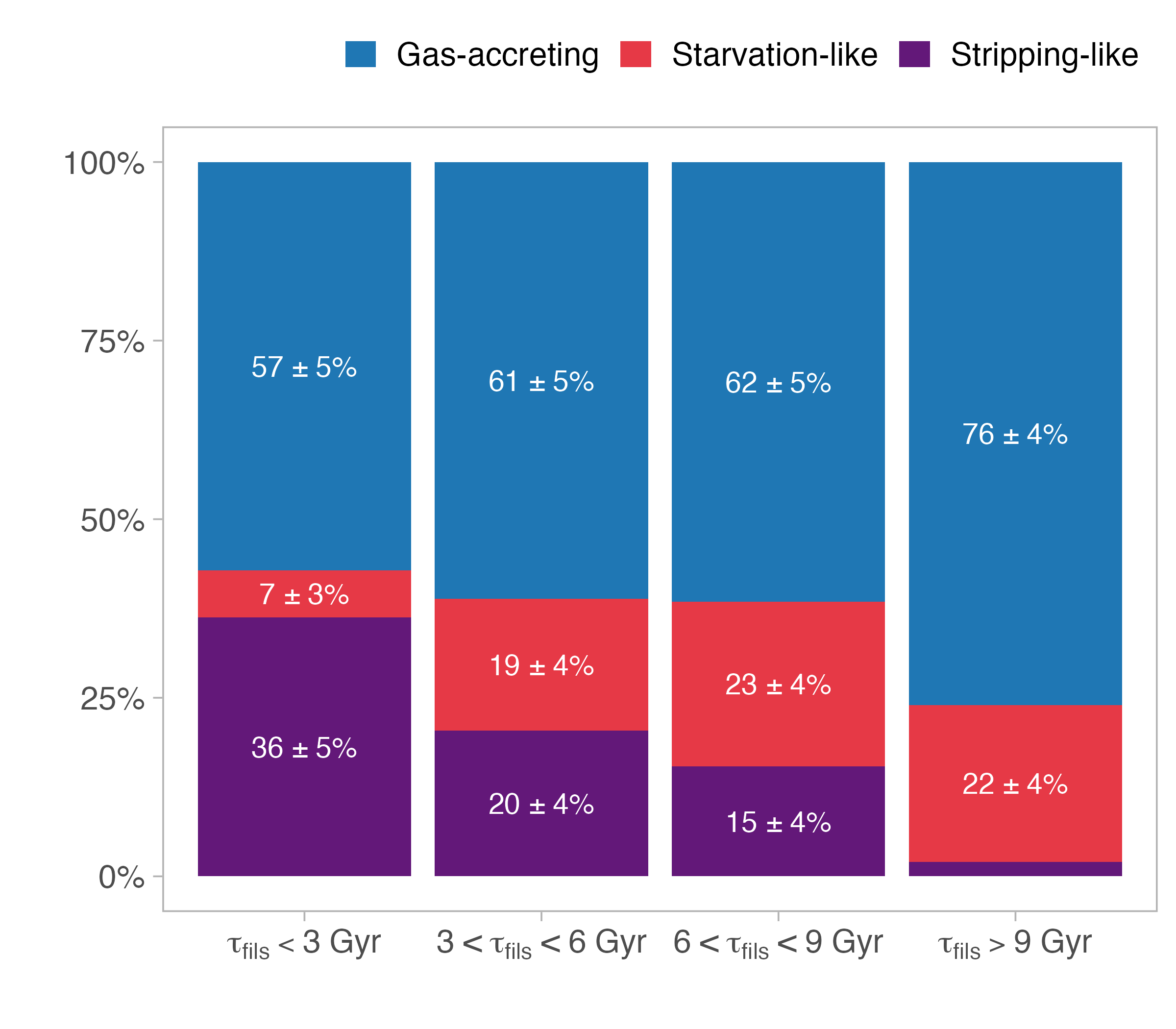}
    \caption{Classification of gas evolution scenarios for filament galaxies, in four bins of filament-infall time. Each bar shows the fractional contribution, with binomial uncertainties indicated inside each segment. }
    \label{fig:starvation_vs_stripping_analysis}
\end{figure}


We apply this procedure to all filament galaxies and estimate the relative occurrence of different cold gas evolution scenarios in bins of infall time in Fig.~\ref{fig:starvation_vs_stripping_analysis}. The dominant phase for low-mass filament galaxies is gas accretion, independent of the infall time. However, a non-negligible fraction of galaxies exhibit a reduction of their cold gas mass after infall. Both the fraction of such systems and the dominant mechanism responsible for the gas depletion depend on the time of infall into the filament. For approximately one quarter of early infallers ($\tau_{\mathrm{fils}} > 9~\mathrm{Gyr}$), the evolution of the cold gas reservoir is consistent with gas consumption by star formation, without evidence for ongoing gas accretion or additional gas removal \textcolor{black}{ and 2\% of the population is in agreement with the stripping scenario}. In Appendix~\ref{app:starvation_vs_stripping_field}, we show that $\sim$90\% of field galaxies classified in the same way are consistent with continued gas accretion. The significantly larger fraction of filament galaxies with declining gas reservoirs indicates the presence of suppressed gas accretion. We argue that this can be explained by a starvation-like scenario driven by the tidal field of the cosmic web, which also leads to the cessation of gas disc growth, as shown in the previous section.
\par
In addition, we identify a population of galaxies whose gas evolution cannot be explained by star formation alone. The fraction of these systems increases towards later infall times, indicating that stripping-like processes become progressively more important. These systems exhibit a stripping-like behaviour, accounting for $15\pm4\%$ and $20\pm4\%$ in the intermediate bins, and reaching $36\pm5\%$ for the most recent infallers ($\tau_{\mathrm{fils}} < 3~\mathrm{Gyr}$). This trend is consistent with the results presented in the previous section. The simultaneous presence of rapid gas removal and gas disc truncation points to cosmic web stripping.

\section{Discussion}
\label{sec:discussion}

We investigate the impact of the filaments on galaxy evolution using the TNG50-1 simulation, focusing on galaxies' gaseous component. Several previous studies report little or no difference between filament and field galaxy populations once stellar or halo mass is controlled for, in terms of their integrated properties such as colours, luminosities, morphologies, and star-formation rates~\citep[e.g.][]{Alpaslan+2015, OKane+2024, Navdha+2025}. 
TNG50-1 predicts that filament galaxies have smaller gas discs than stellar- and halo-mass-matched field galaxies, despite exhibiting similar integrated properties. In Appendix~\ref{sec::assym}, we also show that the morphology of their disc is more asymmetrical than that of field galaxies. \textcolor{black}{ We show that several mechanisms operate within filaments, affecting the gas evolution and accretion of galaxies after infall onto a filament. For low-mass halos in particular, gas accretion is closely associated with the accretion of the dark matter along the filaments of the cosmic web~\citep{Keres+2005, DekelBirnboim2009}, and a modification of the dark matter accretion may therefore accompany a suppression of gas accretion. Even though we do not find a statistically significant difference between the integrated halo assembly of filament and field galaxies, here we investigate whether the dark matter halos of the different filament populations identified in Section~\ref{sec:results} differ beyond their integrated properties, and whether dark matter accretion is affected by filaments. Additionally, the consideration of dark matter halos supports our separation between stripping-like and starvation-like processes presented in Section~\ref{sec:results}, since these two are expected to have a different response of the dark matter halo to these processes. }

\begin{figure}[h!]
    \centering
    \includegraphics[width=1\linewidth]{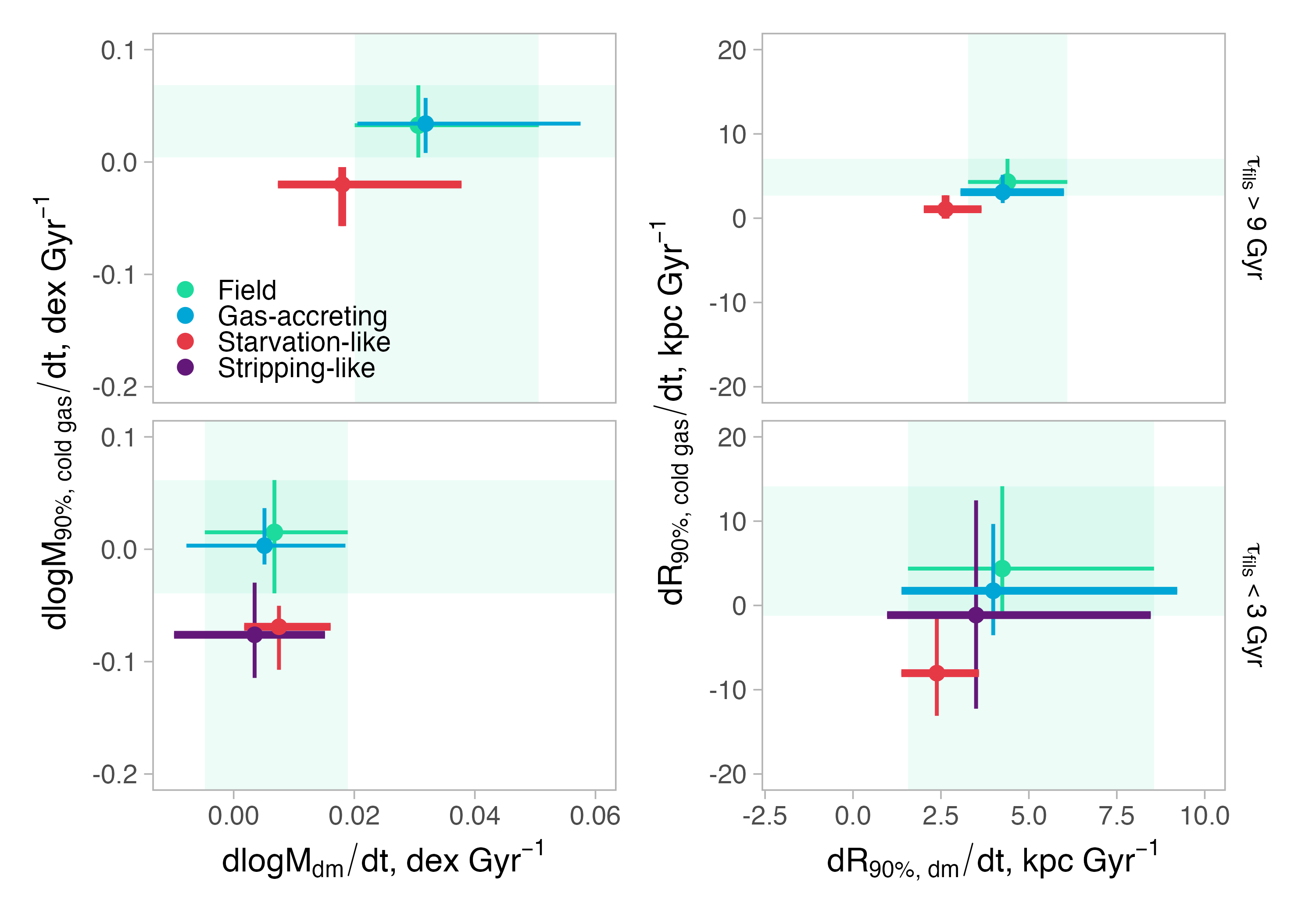}
    \caption{Growth rates of dark matter halo and cold gas disc mass (left) and size (right) for low-mass filament galaxies after infall.
    For each identified gas evolution mechanism after infall, median values with $1\sigma$ uncertainties (16th–84th percentiles) are shown for galaxies with $\tau_{\mathrm{fils}} > 9~\mathrm{Gyr}$ (top) and $\tau_{\mathrm{fils}} < 3~\mathrm{Gyr}$ (bottom). 
    Filament galaxies are compared to their field counterparts, with shaded regions indicating the range of values covered by the field population.}

    \label{fig:mass_and_size_growth_summary_final}
\end{figure}

\par
\textcolor{black}{To quantify changes in dark matter and gas accretion, we estimated the growth rates of dark matter mass, cold gas mass, and their corresponding sizes using the same methodology as described in Section~\ref{subsec:growth_of_disc}. Specifically, for each galaxy, we followed the evolution of the dark matter mass (all dark matter particles gravitationally bound to the subhalo), the dark matter size $R_{90\%,dm}$, the cold gas mass enclosed within $R_{90\%,cold~gas}$, and the cold gas size $R_{90\%,cold~gas}$. The time evolution of each quantity is first smoothed using a LOESS regression. Growth rates were then computed from the smoothed trends using central finite differences. For masses, the procedure was applied to $\log_{10}(M)$, and the resulting growth rates are expressed as $\mathrm{d}\log M/\mathrm{d}t$ in dex per gigayear, while size growth rates are given in kiloparsecs per gigayear. Statistical significance is assessed using two-sample KS tests between filament galaxies and each of the 100 independently matched field realisations. The reported quantity $f(p<0.05)$ corresponds to the fraction of field realisations for which the KS test rejects the null hypothesis with a p-value less than 0.05.}
We focus only on after infall stage and two infall bins, $\tau_{\mathrm{fils}} < 3~\mathrm{Gyr}$ and $\tau_{\mathrm{fils}} > 9~\mathrm{Gyr}$. Fig.~\ref{fig:mass_and_size_growth_summary_final} shows the median growth rates of dark matter and cold gas masses and sizes for these populations, compared to their field counterparts. Recently accreted galaxies have already been shown to exhibit signatures of cosmic web stripping, while for early infallers, the mechanism responsible for the suppression of gas accretion remains less clear. By comparing the evolution of gas and dark matter in these two regimes, we aim to better constrain the physical mechanisms driving this behaviour.
\par
We complement the analysis of mass growth with additional diagnostics of halo dynamics and structure. In particular, we aim to assess whether the suppression of gas accretion in filament galaxies is linked to changes in the dynamical state and assembly of their host dark matter halos.  \citet{Borzyszkowski+2017} showed that low-mass halos affected by tidal fields of the cosmic web exhibit velocity distributions with an enhanced contribution of tangential motions.  Motivated by this, we consider the dark matter velocity anisotropy parameter, $\beta_{\mathrm{DM}} = 1 - \frac{\sigma_t^2}{2\sigma_r^2}$\footnote{\textcolor{black}{$\sigma_r$ is the dispersion of the radial velocity component, and
$\sigma_t^2=\sigma_\theta^2+\sigma_\phi^2$ is the total tangential velocity
dispersion, where $\sigma_\theta$ and $\sigma_\phi$ are the dispersions
of the polar and azimuthal velocity components, respectively.}}, which traces the relative importance of radial and tangential motions. We also quantify halo shape using the minor-to-major axis ratio, $c/a$, as an indicator of anisotropic mass accretion. Values close to 1 correspond to nearly spherical systems, typically associated with dynamically relaxed halos, whereas lower values of $c/a$ indicate increasingly elongated, triaxial structures. Finally, we examine the baryon-to-dark matter mass ratio ($M_{\mathrm{bar}}/M_{\mathrm{DM}}$) and the offset $\rm dx_{SFMR}$ from the star formation rate–stellar mass relation at z=0, to link halo assembly to the present-day gas content and star formation activity of galaxies.

\subsection{Cosmic web tidal fields}
\label{subsec:tidal_fields}

In this section, we focus on galaxies that entered filaments at early cosmic times $\tau_{\mathrm{fils}} > 9~\mathrm{Gyr}$ ago. 
Fig.~\ref{fig:mass_and_size_growth_summary_final} shows that gas-accreting galaxies continue to grow in both dark matter and gas mass at rates comparable to the field, but exhibit suppressed growth of their cold gas discs. Their dark matter and cold gas mass growth rates are 
0.032 and 0.034~$\mathrm{dex~Gyr^{-1}}$, 
consistent with the field values within uncertainties. Statistically significant differences \textcolor{black}{between filaments and each of the field realisations are found } in only 0\% and 1\% of the field realisations. At the same time, their dark matter sizes also grow at rates similar to those in the field 
($\sim4.3~\mathrm{kpc~Gyr^{-1}}$), with significant differences in only 1\% of realisations. 
However, the growth of their cold gas discs is suppressed relative to the field 
(3.1 vs. 4.3~$\mathrm{kpc~Gyr^{-1}}$, although still within uncertainties), 
with statistically significant differences in 100\% of realisations.
\par
\begin{figure}[h!]
    \centering
    \includegraphics[width=1\linewidth]{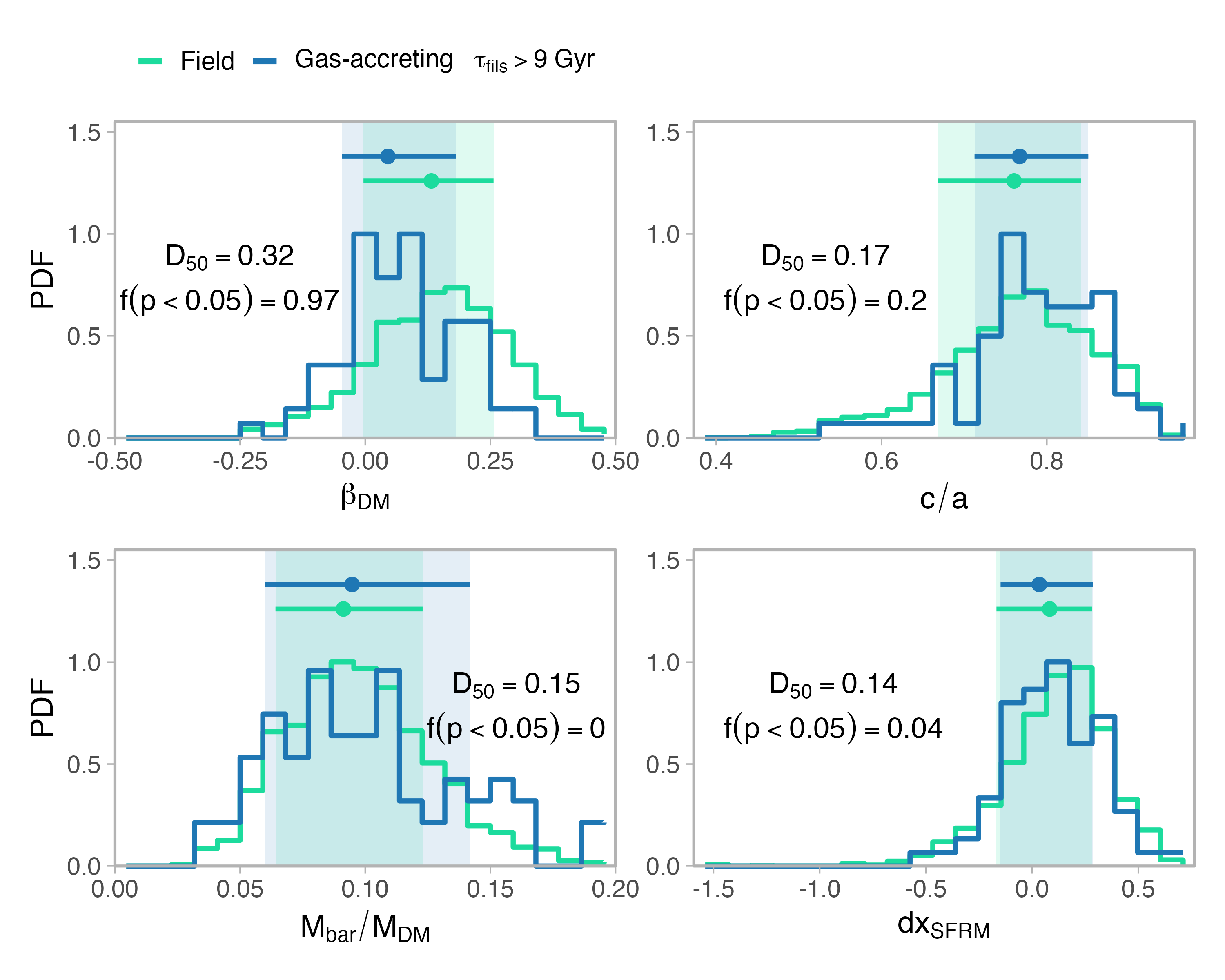}
    \caption{Properties at z=0 of low-mass gas-accreting filament galaxies with $\tau_{\mathrm{fils}} > 9~\mathrm{Gyr}$. Distributions of the dark matter velocity anisotropy parameter $\beta_{\mathrm{DM}}$ (top left), halo shape $c/a$ (top right), baryon-to-dark matter ratio $M_{\mathrm{bar}}/M_{\mathrm{DM}}$ (bottom left), and offset to the star-formation rate-stellar mass main sequence \textcolor{black}{$dx_{\rm SFRM}$} (bottom right) for galaxies that exhibit gas accretion after infall into filaments, compared to stellar- and halo-mass-matched field galaxies. Vertical dashed lines indicate median values, while shaded regions show the 16th–84th percentiles. Each panel reports the median KS statistic $D_{50}$ and the fraction of 100 realisations with $p < 0.05$.    }

    \label{fig:dm_hist_Gas-accreting_taufils9Gyr}
\end{figure}

We explain lower rates of cold gas disc growth with the presence of tidal fields of the cosmic web that affect the geometry of accretion on subhalos. Fig.~\ref{fig:dm_hist_Gas-accreting_taufils9Gyr} shows the velocity anisotropy parameter of dark matter halo and their shape in the top row.  \textcolor{black}{Filament galaxies that continue accreting gas after infall} exhibit a shift of $\beta_{\mathrm{DM}}$ towards negative values, indicating a larger contribution of tangential motions compared to field galaxies, which are characterised by more radially biased orbits. The increased contribution of tangential motions suggests that accretion is no longer dominated by radial infall.  Such a configuration can lead to a more centrally concentrated gas distribution and limit the radial growth of the gas disc, even when the total gas mass and halo growth remain comparable to those of galaxies in the field. In this sense, the smaller gas disc sizes are not driven by a lack of accretion, but by the way in which the accreted material is redistributed within the halo potential~\citep{Mo+1998}. Lastly, we note these galaxies retain halo growth rates, shapes, baryon fractions, and sSFR comparable to their field counterparts.
\par

In contrast, starvation-like systems show reduced growth of both dark matter mass 
(0.018~$\mathrm{dex~Gyr^{-1}}$) and size ($\sim2.6~\mathrm{kpc~Gyr^{-1}}$), 
together with a depletion of their cold gas reservoir driven by star formation 
($-0.020~\mathrm{dex~Gyr^{-1}}$), and almost no growth of the cold gas disc after infall 
($\sim1.1~\mathrm{kpc~Gyr^{-1}}$), with all these differences being statistically significant in 100\% of realisations.

\begin{figure}[h!]
    \centering
    \includegraphics[width=1\linewidth]{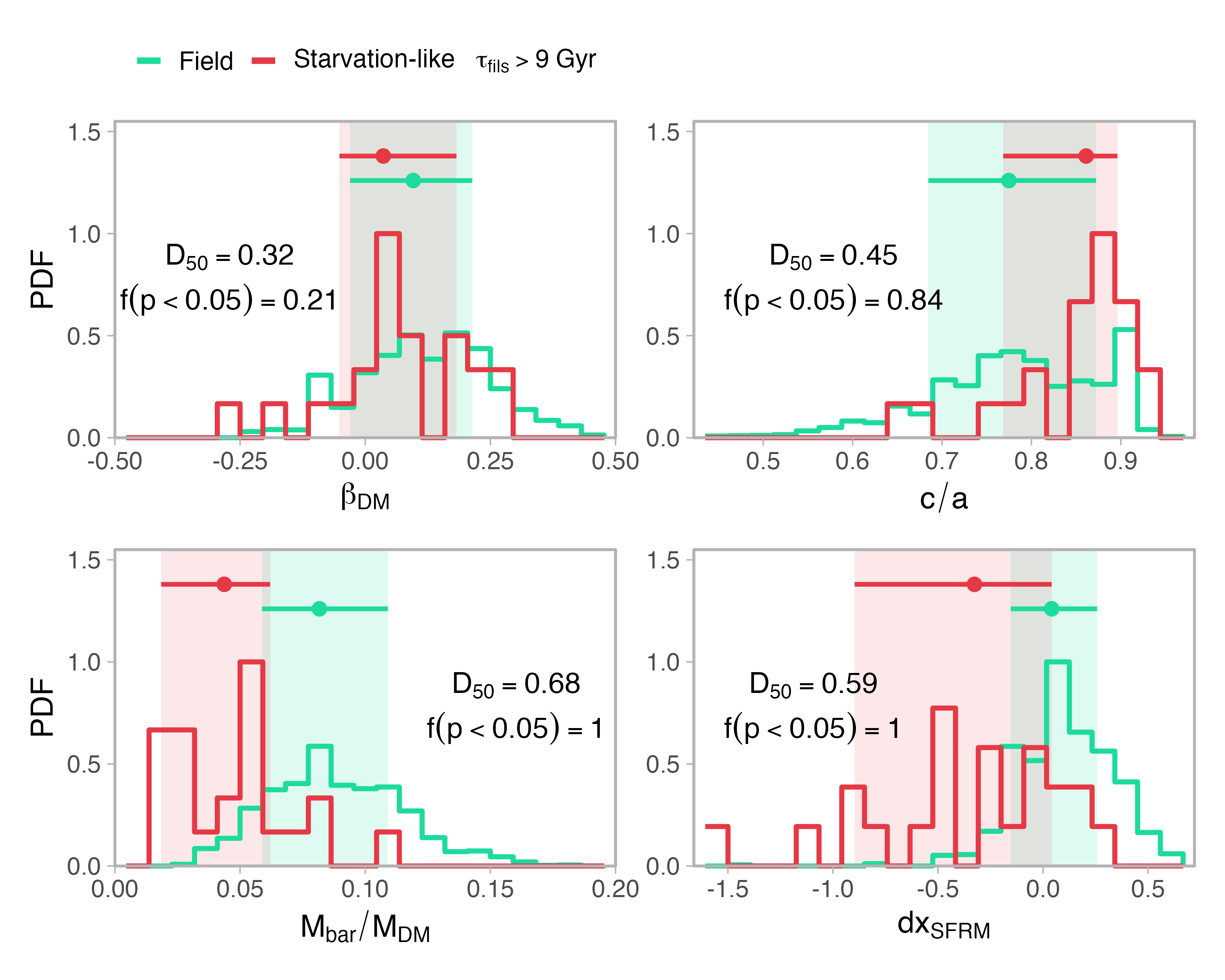}
    \caption{ Same as Fig.~\ref{fig:dm_hist_Gas-accreting_taufils9Gyr} but for starvation-like $\tau_{\mathrm{fils}} > 9~\mathrm{Gyr}$ filament galaxies. }

    \label{fig:dm_hist_Starvation}
\end{figure}

The top left panel of Fig.~\ref{fig:dm_hist_Starvation}  shows that starved filament galaxies and the field have $\beta_{\mathrm{DM}} \approx 0$. The KS test indicates the difference is not statistically significant. 
\textcolor{black}{The} top-right panel shows that starved galaxies reside on average in more spherical halos than field galaxies, with the difference being statistically significant in 84\% of the field-matched realisations. The presence of recent accretion is expected to lead to more elongated halos, while halos with no recent accretion are expected to be characterised by more spherical shapes~\citep{Allgood+2006, Vera-Ciro+2011, Cataldi+2023}. Therefore, we interpret this lack of elongation as evidence of no significant recent dark matter accretion for the starved population. Taken together, these results imply that starved galaxies tend to live in more relaxed halos, consistent with a lower growth rate of the dark matter component.
\par
In addition, starved galaxies exhibit systematically lower baryon-to-dark matter ratios at z=0. We checked that this ratio decreases with time after infall, in contrast to field galaxies, for which it remains approximately constant over time. To quantify this effect, we compare the average growth rates  of the baryon mass (stellar and gas) and dark matter components after infall into filaments, defining
$\Delta g = \left\langle \frac{d\log M_{bar}}{dt} \right\rangle -
\left\langle \frac{d\log M_{\rm DM}}{dt} \right\rangle$.
We find that starvation-like galaxies are systematically shifted towards lower values of $\Delta g$ compared to their field counterparts, with a median $\Delta g = -0.029^{+0.010}_{-0.030}$, while the field distribution is centred around $\Delta g = 0.006^{+0.003}_{-0.004}$.
The KS test confirms that these distributions are statistically distinct: $100\%$ of field realisations have $p < 0.05$. This demonstrates that, after infall into filaments, the growth of baryon mass is more strongly suppressed than that of the dark matter halo, consistent with the expectation that gas is more susceptible to environmental disturbances. This differential impact on baryonic and dark matter components provides evidence that filaments create an imbalance in accretion, preferentially suppressing gas accretion.
\par
Starved filament galaxies show systematically lower $\rm dx_{SFMR}$ compared to field systems, and include a population of quiescent galaxies residing in filaments without any group processing, consistent with a long-term lack of gas supply. We find that the offset reported in Fig.~\ref{fig:dx_evolution} is entirely driven by the starvation-like population: the difference between filament and field galaxies largely disappears once starved galaxies are excluded (in both $6 < \tau_{\mathrm{fils}} < 9$ and $\tau_{\mathrm{fils}} > 9$ bins). This demonstrates that the suppression of star formation in filaments is not a feature of all filament populations, but is instead associated with systems that have ceased to accrete gas.  
\par
Finally, we suggest that starved and gas-accreting populations are affected by the same underlying mechanism, in which the tidal field of the cosmic web modulates the geometry of accretion onto galaxies. The same process acts on both populations but with different efficiency: in some galaxies, it leads to a partial suppression of gas accretion and alters its spatial distribution, primarily affecting the radial growth of gas discs, while in others it results in a more complete shutdown of both gas and dark matter accretion, producing a starvation-like evolution. We note that galaxies with $\tau_{\mathrm{fils}} > 9~\mathrm{Gyr}$ entered filaments close to the peak of cosmic accretion activity. Therefore, any disruption of accretion at these epochs is expected to have long-lasting consequences for both baryonic and dark matter components. 
\par
We speculate that the divergence between starvation-like and gas-accreting systems is determined by a combination of the tidal field of the host filament and the accretion state before infall of the galaxy. Firstly, filaments span a wide range of masses, densities, and tidal field strengths, so their impact on accretion is likely not uniform. Secondly, galaxies with weaker or less coherent accretion flows before infall may be more susceptible to disruption by the filament tidal field, leading to suppressed gas inflow and subsequent starvation. In contrast, galaxies with stronger accretion flows may be able to sustain their inflow despite the influence of the cosmic web, remaining gas-accreting.

\subsection{Cosmic web stripping}

Galaxies that entered filaments at later cosmic times ($\tau_{\mathrm{fils}} < 3~\mathrm{Gyr}$ ago) are expected to be less affected by the tidal field of the cosmic web. This is due to two main reasons. First, these galaxies enter filaments at epochs when accretion is no longer the dominant driver of their growth. Second, these galaxies have spent a relatively short time inside filaments. However, Fig.~\ref{fig:starvation_vs_stripping_analysis} shows that the relative importance of hydrodynamical stripping increases towards z=0, reaching a maximum for $\tau_{\mathrm{fils}} < 3~\mathrm{Gyr}$. This trend can be naturally understood if the efficiency of stripping depends on the density of the intra-filament medium encountered by infalling galaxies. As filaments evolve, they become progressively denser in proper coordinates \citep{Cautun+2014,Martizzi+2019,Bahe+2025},  enhancing the effectiveness of hydrodynamical gas removal processes. For low-mass galaxies, which are characterised by shallower potential wells, this might lead to efficient gas stripping.  In fact, Fig.~\ref{fig:r90_cold_gas_z0} shows that the suppression of cold gas disc sizes is strongest at the low-mass end. We identify two main populations for $\tau_{\mathrm{fils}} < 3~\mathrm{Gyr}$: gas-accreting and stripping-like galaxies. The gas-accreting population exhibits properties very similar to those of field galaxies, appearing largely unaffected by the filaments, as shown in the bottom row of Fig.~\ref{fig:mass_and_size_growth_summary_final}. We interpret this as a consequence of both the short time spent in filaments and the fact that these galaxies do not satisfy the conditions~(filament density or infall velocity) required for efficient stripping. Consistently, their $\beta_{\mathrm{DM}}$, halo shape ($c/a$), baryon-to-dark matter ratio, and $\rm dx_{SFMR}$ are indistinguishable from those of field galaxies. 
\par

\begin{figure}[h!]
    \centering
    \includegraphics[width=1\linewidth]{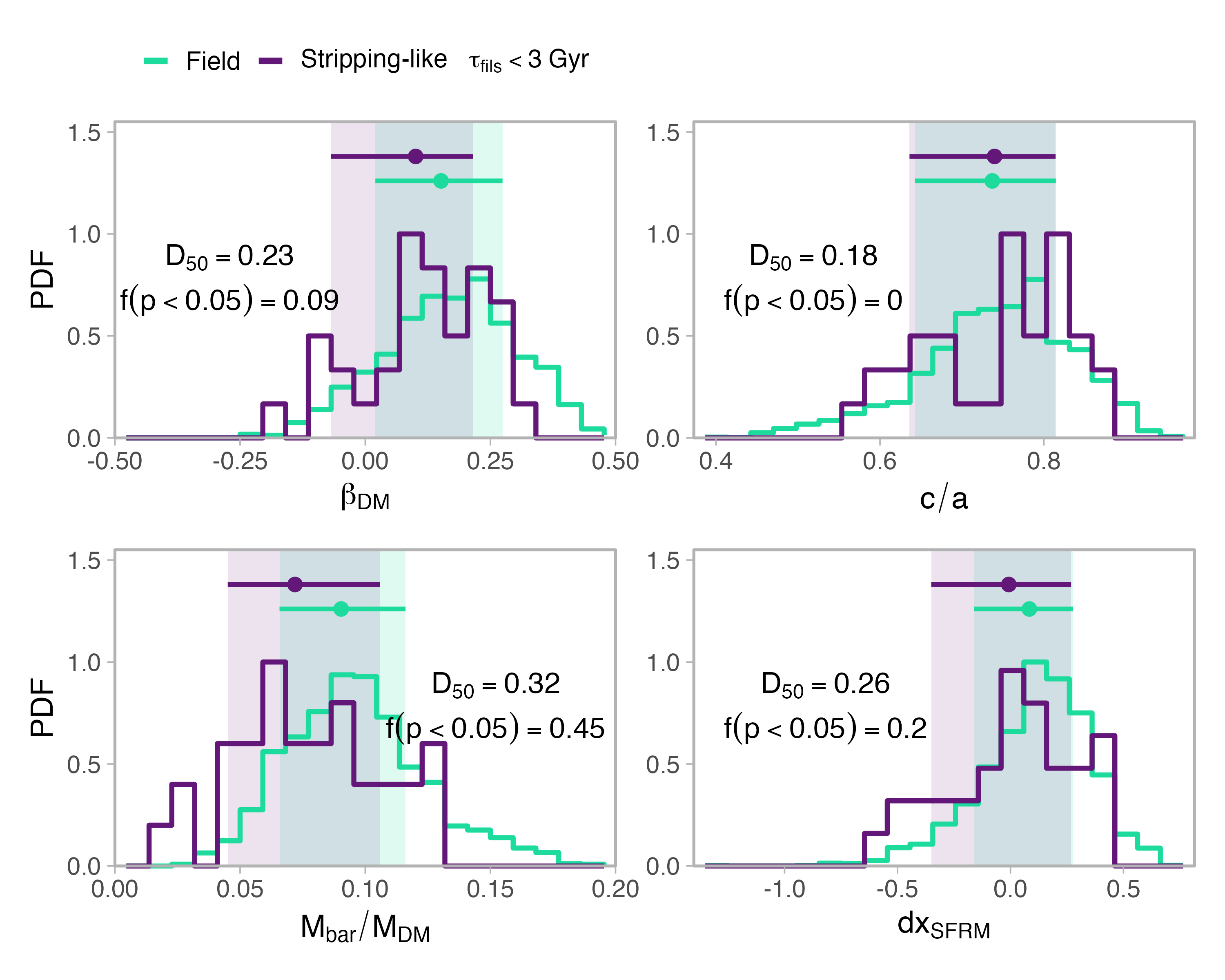}
    \caption{Same as Fig.~\ref{fig:dm_hist_Gas-accreting_taufils9Gyr}, but for filament galaxies with $\tau_{\mathrm{fils}} < 3~\mathrm{Gyr}$ that exhibit stripping-like mechanism after infall. }

    \label{fig:dm_hist_Stripping-like_taufils3Gyr}
\end{figure}


Fig.~\ref{fig:mass_and_size_growth_summary_final} shows that galaxies identified as the stripped population exhibit a significant reduction in cold gas mass and disc size, while their dark matter component remains unaffected, revealing the hydrodynamical nature expected for cosmic web stripping. Fig.~\ref{fig:dm_hist_Stripping-like_taufils3Gyr} further demonstrates that the dynamical state and shape of the dark matter halo are preserved, with the main effect being a reduction in the baryon-to-dark matter fraction bound to the galaxy. This behaviour is consistent with a stripping–like mechanism operating within filaments, where the interaction between galaxies and the intra-filament medium removes loosely bound cold gas from the outer regions, leading to a truncation of the discs as shown in Fig.~\ref{fig:size_evolution_anal}. The fact that star-forming regions remain unaffected (Fig.~\ref{fig:r90_sfr_gas_z0}) while the outer cold gas disc is truncated indicates that stripping proceeds in an outside-in fashion, preferentially removing low-density gas at large radii, while leaving unaffected the denser gas where star formation takes place. The enhanced asymmetry of stripped galaxies \textcolor{black}{(Fig.~\ref{fig:assymentry_as_stm})} further supports this interpretation, as asymmetric gas removal naturally produces lopsided gas morphologies.

\subsection{Geometry of infall}

The geometry of infall, and in particular the orientation of the galaxy with respect to the filament at the time of accretion, might play a role in determining which mechanism operates within filaments. In fact, the orientation of infall may alter or partially suppress the inflow of material along the filament, potentially leading to starvation-like evolution. Similarly, stripping-like processes are expected to depend on the relative velocity and orientation of a galaxy with respect to the surrounding medium, as established in the context of ram-pressure stripping in groups and clusters (e.g.~\citealt{Roediger+2006, Bahe+2013, Steinhauser+2016}). 
\par
\begin{figure}
    \centering
    \includegraphics[width=1\linewidth]{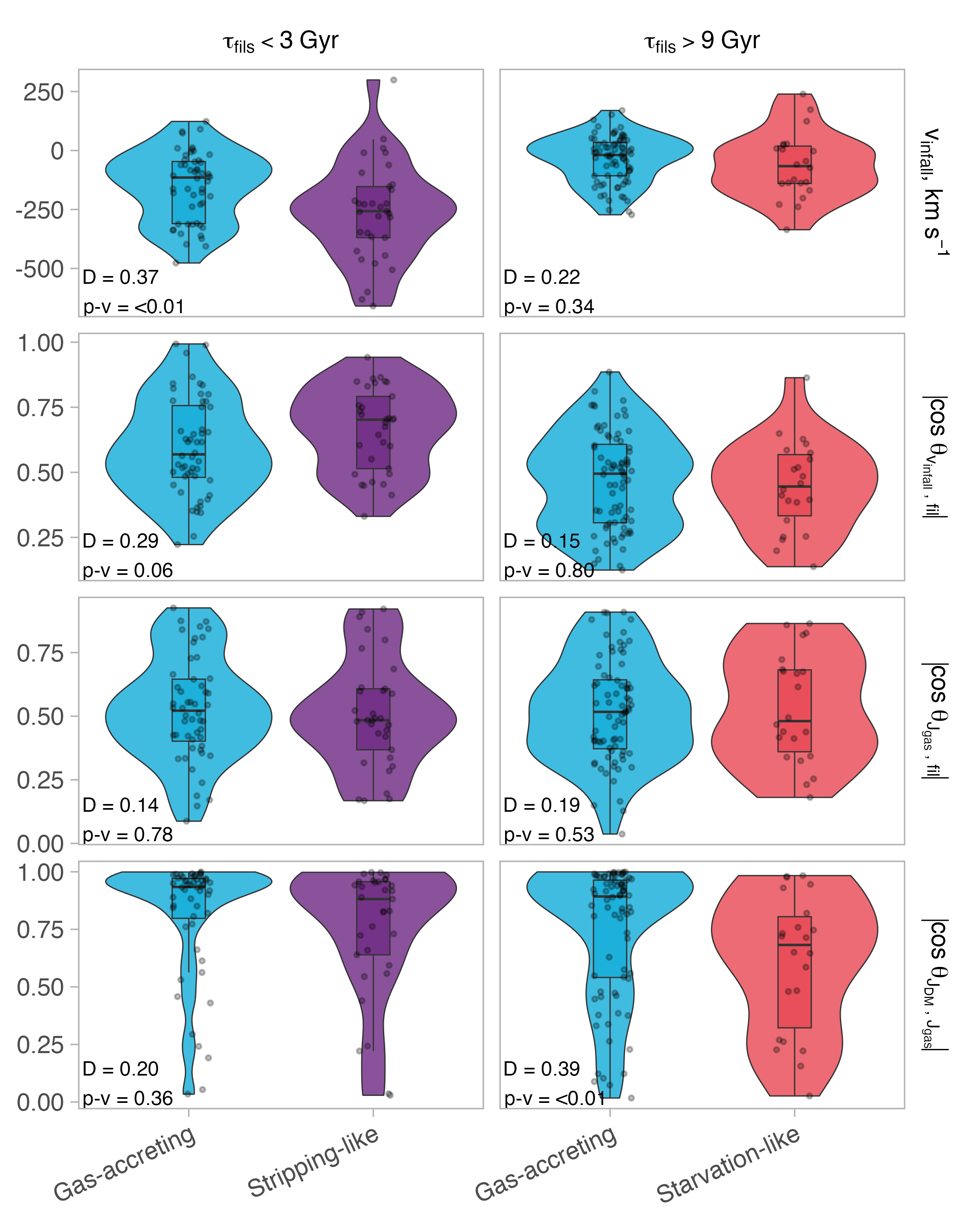}
    \caption{ Geometry of infall into filaments for low-mass galaxies with $\tau_{\mathrm{fils}} < 3~\mathrm{Gyr}$ ago and $\tau_{\mathrm{fils}} > 9~\mathrm{Gyr}$ ago. Distributions are shown for the dominant channels of gas evolution identified in Fig.~\ref{fig:starvation_vs_stripping_analysis}. The different rows display the distributions of infall velocity (top), the orientation of the infall trajectory relative to the filament (second row), the alignment of the gas angular momentum with the filament (third row), and the relative alignment between the gas and dark matter angular momenta (bottom). The bottom left corner of each panel shows the results of the KS test between the compared groups (D statistic and p-value).}

    \label{fig:geometry_of_infall}
\end{figure}

To characterise the geometry of infall, we define the following quantities for each galaxy at the moment of infall:

\begin{itemize}

    \item the infall velocity, $\rm v_{\rm infall} = ~\mathbf{v}\cdot\hat{\mathbf{r}}_{\rm fil}$ (km~s$^{-1}$), \textcolor{black}{where $\hat{\mathbf r}_{\rm fil}$ is the unit vector pointing from the
nearest point on the filament spine towards the galaxy}, defined as the radial component of the galaxy velocity with respect to the filament axis (top row of Fig.~\ref{fig:geometry_of_infall});
    
    \item the orientation of the infall trajectory, defined as $\rm |\cos \theta_{v_{infall},fil}| = \langle |\hat{\mathbf v}\cdot\hat{\mathbf r}_{\rm fil}| \rangle$, which measures the angle between the galaxy velocity vector and the radial direction towards the filament axis. \textcolor{black}{$\rm |\cos \theta_{v_{infall},fil}|=1$ corresponds to radial motion, and $\rm |\cos \theta_{v_{infall},fil}|=0$ }to motion along the filament (second row of Fig.~\ref{fig:geometry_of_infall});
    
    \item the orientation of the gas angular momentum with respect to the filament, $\rm |\cos \theta_{J_{\rm gas}, fil}| = \langle |\hat{\mathbf J}_{\rm gas}\cdot\hat{\mathbf t}_{\rm fil}| \rangle$, \textcolor{black}{where $\hat{\mathbf t}_{\rm fil}$ is the unit vector tangent to the
    filament spine at the point nearest to the galaxy}, which describes the alignment between the gas angular momentum and the filament axis. \textcolor{black}{$\rm |\cos \theta_{J_{\rm gas}, fil}|=1$ indicates alignment, and $\rm |\cos \theta_{J_{\rm gas}, fil}|=0$} indicates a perpendicular configuration (third row of Fig.~\ref{fig:geometry_of_infall});
    
    \item the relative orientation between the gas and dark matter angular momentum, $\rm |\cos \theta_{J_{\rm DM},J_{\rm gas}}| = \langle |\hat{\mathbf J}_{\rm DM}\cdot\hat{\mathbf J}_{\rm gas}| \rangle$, which quantifies the alignment between the gas and dark matter components (bottom row of Fig.~\ref{fig:geometry_of_infall});

\end{itemize}

Each panel in Fig.~\ref{fig:geometry_of_infall} shows the distribution of these quantities for the dominant mechanism of evolution inside filaments (gas-accreting, stripping-like, and starvation-like populations), separated by infall time. All listed quantities are computed for each of the ten filament realisations at a given snapshot, and the median value is reported for each galaxy.

\par
The first column of Fig.~\ref{fig:geometry_of_infall} shows the infall geometry of galaxies that entered filaments recently, $\tau_{\mathrm{fils}} < 3~\mathrm{Gyr}$ ago.  Galaxies that remain gas-accreting after infall and those that exhibit stripping-like evolution show different distributions of $\rm v_{\rm infall}$ (top left panel), both visually and according to the KS test. In particular, stripping-like galaxies have systematically higher infall velocities than gas-accreting systems: the median infall velocity is $115 \pm 158,\mathrm{km~s^{-1}}$ for the gas-accreting population, compared to $259 \pm 194,\mathrm{km~s^{-1}}$ for stripping-like galaxies. This suggests that infall velocity, in combination with the density of the intra-filament medium, is a key factor governing whether a galaxy undergoes gas stripping inside filaments, and that this becomes effective only for sufficiently large velocities.  The orientation of the infall trajectory (second row of Fig.~\ref{fig:geometry_of_infall}) also appears to play a role. At infall, stripping-like galaxies show a distribution of $\rm |\cos \theta_{v_{\rm infall},fil}|$ that is shifted towards more radial infall with respect to gas-accreting galaxies, which exhibit a more isotropic distribution. However, the KS test does not reveal a statistically significant difference between these distributions, \textcolor{black}{confirming that the infall trajectories of stripped galaxies are not distinguishable from those of the other populations.} The last two rows for $\tau_{\mathrm{fils}} < 3~\mathrm{Gyr}$ ago do not reveal any significant differences between the gas-accreting and stripping-like populations:  neither the orientation of the gas disc nor the coherence between gas motion and dark matter plays a role in determining the fate of galaxies after infall. This behaviour is consistent with a picture in which gas stripping inside filaments operates analogously to ram-pressure stripping in clusters and groups, with the controlling parameters being the relative velocity and ambient density, rather than the internal geometry of the galaxy.
\par
The second column of Fig.~\ref{fig:geometry_of_infall} shows the infall geometry of early infallers  ($\tau_{\mathrm{fils}} > 9~\mathrm{Gyr}$ ago). We find that gas-accreting and starvation-like galaxies have similar distributions of infall velocities, with no evidence that higher infall velocities lead to enhanced disruption of accretion flows. Similarly, neither the orientation of the infall trajectory nor the orientation of the gas disc shows any significant difference between the two populations. In contrast, the last row of Fig.~\ref{fig:geometry_of_infall} shows that galaxies that undergo starvation already exhibit a misalignment between the angular momentum of dark matter and gas at the time of infall. Such a misalignment may indicate less coherent gas accretion, making it more susceptible to disruption by tidal fields, which may then suppress further accretion. This misalignment could itself be a consequence of tidal interactions rather than the cause of already started suppression of gas accretion. We note that by z=0 both populations exhibit aligned dark matter and gas angular momenta, as starvation-like systems relax over time, while gas-accreting galaxies do not develop significant misalignment.

\subsection{Caveats}
\label{subsec::caveats}

We have argued that the results presented above point to a consistent picture of the impact of the cosmic web tidal field and cosmic web stripping. However, it is important to consider a few caveats and modelling limitations that may influence our measurements and interpretation.
\par
We verified that these differences are not driven by thermal effects. 
In particular, galaxies in filaments do not exhibit systematically different gas temperature profiles compared to their field counterparts (see Appendix~\ref{app:temp_profile}). 
\par
We note that the classification adopted in Sec.~\ref{subsec:mass_classification} is model-dependent and relies on the assumed reference star-formation rates of field galaxies. While we adopt the median field relation, the field population spans a broad range of star-formation rates, and adopting a higher reference level would shift most galaxies into the starvation-like regime, effectively removing the stripping-like category. Despite this, we argue that the stripping-like population is physically well motivated. Compared to starvation, these systems exhibit faster gas removal than expected from star formation alone, truncated gas discs rather than halted growth, and no corresponding changes in dark matter halo properties. In contrast, starvation-like galaxies show suppressed accretion in both gas and dark matter. 
\par
Our classification is based on the net change in cold gas mass between infall and z=0, and therefore does not capture more complex evolutionary pathways. For example, a galaxy may initially accrete gas after infall and later experience stripping. In such cases, the system may be classified as starvation-like despite having undergone stripping at later times. While more sophisticated classifications based on the time that galaxies spend in a type of gas loss or accretion could be adopted, they are less straightforward to interpret. Nevertheless, we have verified that all conclusions presented beyond Sec.~\ref{subsec:mass_classification} remain robust under alternative classification schemes, with differences affecting only the relative fractions shown in Fig.~\ref{fig:starvation_vs_stripping_analysis}.
\par
We note that our analysis is subject to the limited resolution and volume of TNG50. In particular, the sample size is modest, with of order $\sim100$ galaxies in each $\tau_{\mathrm{fils}}$ bin, and significantly fewer in some subsamples (e.g. only $22$ starvation-like galaxies for $\tau_{\mathrm{fils}} > 9~\mathrm{Gyr}$). 
\par
Finally, throughout the paper, we argue that we detect signatures of cosmic web stripping and starvation acting on galaxies within filaments beyond classical group pre-processing. In our main analysis, we exclude galaxies identified as non-central members of a halo for at least two consecutive snapshots. However, environmental effects from massive halos can extend well beyond the virial radius, out to $\sim 2$--$3~R_{\rm vir}$ \citep{Behroozi+2014}. Moreover, filaments are populated by groups and massive halos \citep{Tempel+2014_perls}, and galaxies in filaments are more likely than field galaxies to experience processing by nearby halos  \textcolor{black}{or interact with other galaxies~(e.g. fly-bys), and including more frequent merging~\citep{Zakharova_envI}. These processes can also affect our galaxies beyond the discussed cosmic web stripping or cosmic web tidal field. }
To assess the extent to which our stripping-like galaxies are affected by neighbouring halos \textcolor{black}{or galaxies} embedded within filaments, we ran the following test. We measured their distance, at each snapshot, to the nearest dark matter halo with $\log(M_{\rm halo}/M_{sun}) > 10$, roughly corresponding to a system capable of hosting a galaxy with $\log(M_{star}/M_{sun})\approx 8$, \textcolor{black}{allowing us to account not only for the impact of close groups or clusters but also for galaxy-galaxy interactions}. We then determined how many galaxies that exhibit stripping-like or starvation-like evolution have ever approached such a halo within $3~R_{\rm vir}$ over their lifetime. This criterion is very conservative: we do not expect every halo with $\log(M_{\rm halo}/M_{sun}) > 10$ to influence galaxies out to $3~R_{\rm vir}$ with the same efficiency as clusters in \cite{Behroozi+2014}, but it provides a stringent upper limit on the possible role of halo-driven processing. We find that about $50\%$ of our stripping-like and starvation-like filament galaxies have ever come within $3~R_{\rm vir}$ of a halo with $\log(M_{\rm halo}/M_{sun}) > 10$. Therefore, at least half (but likely more than this) of the filament galaxies that experienced stripping-like evolution were not affected by neighbouring halos. If we adopt a less stringent distance threshold of $2~R_{\rm vir}$, the fraction of galaxies that are affected by the filament-driven mechanism but not by a halo rises to $\sim 70\%$. We stress that this estimate is approximate. Galaxies inside filaments tend to flow toward groups and clusters, especially at late times, so close passages near massive halos are expected even if the physical mechanism driving gas evolution operated earlier and was primarily filamentary in origin. Our simple test does not explicitly account for this timing information.  \textcolor{black}{We note that we do not account for interaction with galaxies or halos of lower mass and merger events between low-mass galaxies. The latter are expected to be negligible because the stellar mass assembly histories of filament and field low-mass galaxies are statistically indistinguishable, which would not be the case if mergers contributed differentially to gas removal between the two populations.}

\section{Summary and conclusions}
\label{sec:conclusion}

 In this paper, we use the TNG50 hydrodynamical simulation to study the \textcolor{black}{mechanisms that affect the evolution of low-mass~($8 \le \log(M_{star}/M_{sun}) \le 10$) galaxies in }filaments. 
 \textcolor{black}{We compared filament galaxies with field galaxies of the same stellar and halo mass, and to isolate filament-driven mechanisms, we restricted both samples to galaxies that remained centrals throughout their evolution, thereby minimising contamination from group- and cluster-specific impact. }
 No significant difference is found between the \textcolor{black}{filament and field} populations in terms of stellar or halo mass growth and quenched fraction. On the other hand, we demonstrate that filament galaxies exhibit consistently smaller cold gas discs at both $R_{90\%,\mathrm{cold~gas}}$ and $R_{50\%,\mathrm{cold~gas}}$ radii, enhanced asymmetries in their cold gas discs, and, in some cases, reduced star formation within filaments.
 \textcolor{black}{We identify several mechanisms operating within filaments and show that their relative importance depends on the cosmic epoch at which galaxies infall into filaments.}
 The infall time determines both the duration of the filamentary influence on a galaxy and the physical conditions it encounters, which evolve over cosmic time.  
 Below, we summarise the key mechanisms identified in this work:

\begin{itemize}

    \item Galaxies that fall into filaments at early cosmic times, $\tau_{\mathrm{fils}} > 9~\mathrm{Gyr}$ ago, are affected by the cosmic web tidal field, which disrupts the accretion of both gas and dark matter. \textcolor{black}{This mechanism is particularly important for early infallers, as they enter filaments during an epoch of active mass assembly, when changes in the accretion process can strongly affect their subsequent evolution. In addition, these galaxies spend most of their lifetime within filaments, allowing the filamentary influence to act over several gigayears.}
    We identify two populations of early infallers affected by the tidal field of the cosmic web. In the first case, galaxies continue gas and dark matter accretion at rates comparable to those in the field. \textcolor{black}{However, such galaxies have smaller gas discs even though they have the same cold gas reservoir. } 
    \textcolor{black}{ We found that these galaxies are characterised by a more tangentially biased velocity anisotropy in their dark matter halos. We interpret this as evidence that the tidal field of the cosmic web modifies the geometry of accretion, resulting in a larger contribution of tangential motions relative to radial infall. We suggest that the larger contribution of tangential motions limits the radial growth of the gas distribution, as tangential trajectories reach smaller radial distances than predominantly radial infall. This may explain the systematically smaller gas discs of these galaxies.}
    \textcolor{black}{In contrast, the second population of early infallers exhibits strongly suppressed growth of both the gas and dark matter components. We interpret this as a regime in which the cosmic web tidal field disrupts accretion, reducing the supply of fresh gas and leading to starvation. }
    \textcolor{black}{The sustained suppression of gas accretion gradually depletes the cold gas reservoir. We identify a population of quiescent galaxies among these starvation-like systems. These results suggest that cosmic web tidal fields can contribute to the quenching of low-mass galaxies}.

    \item Galaxies that entered filaments more recently ($\tau_{\mathrm{fils}} < 3~\mathrm{Gyr}$) \textcolor{black}{can also be broadly divided into two populations by the filament impact. The first one shows} evidence of rapid gas loss, truncation of their gas discs, and enhanced asymmetries, while their dark matter halos remain largely unaffected. These signatures are consistent with hydrodynamical cosmic web stripping. This process is also mass-dependent and predominantly affects galaxies at a lower mass end, which are more susceptible to gas removal due to their shallower potential wells. A key factor regulating this mechanism is the infall velocity into the filament, which determines the strength of the interaction with the intra-filament medium similarly to ram-pressure stripping in clusters. Its increasing importance towards later times further suggests a dependence on the evolving properties of filaments, becoming more efficient in denser environments.  \textcolor{black}{The second population consists of galaxies for which we detect no significant filament-driven changes in the properties considered. Their apparent lack of response may be due to the relatively short time spent within filaments and, possibly, to their higher stellar masses, which make them less susceptible to stripping. }

\end{itemize}

Our results demonstrate that filaments play a non-negligible role in shaping the evolution of low-mass galaxies. Within the TNG50-1 model, there is no evidence of enhanced gas accretion in filaments for galaxies in the mass bin considered. Rather, our results point to a systematic suppression or redistribution of cold gas, leading to statistically smaller and more asymmetric gas discs across the filament population, while the star-forming regions remain largely unaffected.

\begin{acknowledgements}
\textcolor{black}{We thank Simon White for the constructive feedback that improved this work. We thank an anonymous referee for careful reading and thoughtful comments that helped to improve the clarity of the manuscript. Generative AI models were used to improve the language and clarity of the text.} 
\end{acknowledgements}

%

  \bibliographystyle{aa} 
  \bibliography{main}

\begin{thebibliography}{92}
\expandafter\ifx\csname natexlab\endcsname\relax\def\natexlab#1{#1}\fi

\bibitem[{{Allgood} {et~al.}(2006){Allgood}, {Flores}, {Primack}, {Kravtsov}, {Wechsler}, {Faltenbacher}, \& {Bullock}}]{Allgood+2006}
{Allgood}, B., {Flores}, R.~A., {Primack}, J.~R., {et~al.} 2006, \mnras, 367, 1781

\bibitem[{{Alpaslan} {et~al.}(2015){Alpaslan}, {Driver}, {Robotham}, {Obreschkow}, {Andrae}, {Cluver}, {Kelvin}, {Lange}, {Owers}, {Taylor}, {Andrews}, {Bamford}, {Bland-Hawthorn}, {Brough}, {Brown}, {Colless}, {Davies}, {Eardley}, {Grootes}, {Hopkins}, {Kennedy}, {Liske}, {Lara-L{\'o}pez}, {L{\'o}pez-S{\'a}nchez}, {Loveday}, {Madore}, {Mahajan}, {Meyer}, {Moffett}, {Norberg}, {Penny}, {Pimbblet}, {Popescu}, {Seibert}, \& {Tuffs}}]{Alpaslan+2015}
{Alpaslan}, M., {Driver}, S., {Robotham}, A. S.~G., {et~al.} 2015, \mnras, 451, 3249

\bibitem[{{Aragon Calvo} {et~al.}(2019){Aragon Calvo}, {Neyrinck}, \& {Silk}}]{Aragon_Calvo+2019}
{Aragon Calvo}, M.~A., {Neyrinck}, M.~C., \& {Silk}, J. 2019, The Open Journal of Astrophysics, 2, 7

\bibitem[{{Bah{\'e}} \& {Jablonka}(2025)}]{Bahe+2025}
{Bah{\'e}}, Y.~M. \& {Jablonka}, P. 2025, \aap, 702, A145

\bibitem[{{Bah{\'e}} {et~al.}(2013){Bah{\'e}}, {McCarthy}, {Balogh}, \& {Font}}]{Bahe+2013}
{Bah{\'e}}, Y.~M., {McCarthy}, I.~G., {Balogh}, M.~L., \& {Font}, A.~S. 2013, \mnras, 430, 3017

\bibitem[{{Barsanti} {et~al.}(2022){Barsanti}, {Colless}, {Welker}, {Oh}, {Casura}, {Bryant}, {Croom}, {D'Eugenio}, {Lawrence}, {Richards}, \& {van de Sande}}]{Barsanti+2022}
{Barsanti}, S., {Colless}, M., {Welker}, C., {et~al.} 2022, \mnras, 516, 3569

\bibitem[{{Barsanti} {et~al.}(2025){Barsanti}, {Croom}, {Colless}, {Bland-Hawthorn}, {Brough}, {Bryant}, {Lorente}, {Oh}, {Santucci}, {Sweet}, {Sande van de}, \& {Welker}}]{Barsanti+2025}
{Barsanti}, S., {Croom}, S.~M., {Colless}, M., {et~al.} 2025, \mnras, 538, 2660

\bibitem[{{Barsanti} {et~al.}(2026){Barsanti}, {Wang}, {Colless}, {Liu}, {Bulbul}, {Owers}, {Croom}, {Vulcani}, {Bryant}, {Mai}, {Oh}, {Ristea}, {Sweet}, \& {van de Sande}}]{Barsanti+2026}
{Barsanti}, S., {Wang}, D., {Colless}, M., {et~al.} 2026, \mnras, 549, stag862

\bibitem[{{Behroozi} {et~al.}(2014){Behroozi}, {Wechsler}, {Lu}, {Hahn}, {Busha}, {Klypin}, \& {Primack}}]{Behroozi+2014}
{Behroozi}, P.~S., {Wechsler}, R.~H., {Lu}, Y., {et~al.} 2014, \apj, 787, 156

\bibitem[{{Benavides} {et~al.}(2025){Benavides}, {Navarro}, {Sales}, {P{\'e}rez}, \& {Bidaran}}]{Benavides+2025}
{Benavides}, J.~A., {Navarro}, J.~F., {Sales}, L.~V., {P{\'e}rez}, I., \& {Bidaran}, B. 2025, \apj, 985, 86

\bibitem[{{Ben{\'\i}tez-Llambay} {et~al.}(2013){Ben{\'\i}tez-Llambay}, {Navarro}, {Abadi}, {Gottl{\"o}ber}, {Yepes}, {Hoffman}, \& {Steinmetz}}]{Benitez-Llambay+2013}
{Ben{\'\i}tez-Llambay}, A., {Navarro}, J.~F., {Abadi}, M.~G., {et~al.} 2013, \apjl, 763, L41

\bibitem[{{Bilimogga} {et~al.}(2026){Bilimogga}, {Busekool}, {Verheijen}, \& {van der Hulst}}]{Bilimogga+2026}
{Bilimogga}, P.~V., {Busekool}, E., {Verheijen}, M.~A.~W., \& {van der Hulst}, J.~M. 2026, \aap, 710, A159

\bibitem[{{Bond} {et~al.}(1996){Bond}, {Kofman}, \& {Pogosyan}}]{Bond+1996}
{Bond}, J.~R., {Kofman}, L., \& {Pogosyan}, D. 1996, \nat, 380, 603

\bibitem[{{Borzyszkowski} {et~al.}(2017){Borzyszkowski}, {Porciani}, {Romano-D{\'\i}az}, \& {Garaldi}}]{Borzyszkowski+2017}
{Borzyszkowski}, M., {Porciani}, C., {Romano-D{\'\i}az}, E., \& {Garaldi}, E. 2017, \mnras, 469, 594

\bibitem[{{Bournaud} {et~al.}(2005){Bournaud}, {Combes}, {Jog}, \& {Puerari}}]{Bournaud+2005}
{Bournaud}, F., {Combes}, F., {Jog}, C.~J., \& {Puerari}, I. 2005, \aap, 438, 507

\bibitem[{{Cadiou} {et~al.}(2020){Cadiou}, {Pichon}, {Codis}, {Musso}, {Pogosyan}, {Dubois}, {Cardoso}, \& {Prunet}}]{Cadiou+2020}
{Cadiou}, C., {Pichon}, C., {Codis}, S., {et~al.} 2020, \mnras, 496, 4787

\bibitem[{{Castignani} {et~al.}(2022{\natexlab{a}}){Castignani}, {Combes}, {Jablonka}, {Finn}, {Rudnick}, {Vulcani}, {Desai}, {Zaritsky}, \& {Salom{\'e}}}]{Castignani+2022_gas}
{Castignani}, G., {Combes}, F., {Jablonka}, P., {et~al.} 2022{\natexlab{a}}, \aap, 657, A9

\bibitem[{{Castignani} {et~al.}(2022{\natexlab{b}}){Castignani}, {Vulcani}, {Finn}, {Combes}, {Jablonka}, {Rudnick}, {Zaritsky}, {Whalen}, {Conger}, {De Lucia}, {Desai}, {Koopmann}, {Moustakas}, {Norman}, \& {Townsend}}]{Castignani+2022_catalogue}
{Castignani}, G., {Vulcani}, B., {Finn}, R.~A., {et~al.} 2022{\natexlab{b}}, \apjs, 259, 43

\bibitem[{{Cataldi} {et~al.}(2023){Cataldi}, {Pedrosa}, {Tissera}, {Artale}, {Padilla}, {Dominguez-Tenreiro}, {Bignone}, {Gonzalez}, \& {Pellizza}}]{Cataldi+2023}
{Cataldi}, P., {Pedrosa}, S.~E., {Tissera}, P.~B., {et~al.} 2023, \mnras, 523, 1919

\bibitem[{{Cautun} {et~al.}(2014){Cautun}, {van de Weygaert}, {Jones}, \& {Frenk}}]{Cautun+2014}
{Cautun}, M., {van de Weygaert}, R., {Jones}, B. J.~T., \& {Frenk}, C.~S. 2014, \mnras, 441, 2923

\bibitem[{{Chen} {et~al.}(2017){Chen}, {Ho}, {Mandelbaum}, {Bahcall}, {Brownstein}, {Freeman}, {Genovese}, {Schneider}, \& {Wasserman}}]{Chen+2017}
{Chen}, Y.-C., {Ho}, S., {Mandelbaum}, R., {et~al.} 2017, \mnras, 466, 1880

\bibitem[{{Codis} {et~al.}(2018){Codis}, {Jindal}, {Chisari}, {Vibert}, {Dubois}, {Pichon}, \& {Devriendt}}]{Codis_orient}
{Codis}, S., {Jindal}, A., {Chisari}, N.~E., {et~al.} 2018, \mnras, 481, 4753

\bibitem[{{Conger} {et~al.}(2025){Conger}, {Rudnick}, {Finn}, {Castignani}, {Moustakas}, {Vulcani}, {Zakharova}, {Xie}, {Combes}, {Jablonka}, {Bah{\'e}}, {De Lucia}, {Desai}, {Koopmann}, {Norman}, {Townsend}, \& {Zaritsky}}]{Conger+2025}
{Conger}, K., {Rudnick}, G., {Finn}, R.~A., {et~al.} 2025, \apj, 978, 113

\bibitem[{{Crone Odekon} {et~al.}(2018){Crone Odekon}, {Hallenbeck}, {Haynes}, {Koopmann}, {Phi}, \& {Wolfe}}]{odekon+2018}
{Crone Odekon}, M., {Hallenbeck}, G., {Haynes}, M.~P., {et~al.} 2018, \apj, 852, 142

\bibitem[{{De Lucia} {et~al.}(2024){De Lucia}, {Fontanot}, {Xie}, \& {Hirschmann}}]{De_Lucia+2024}
{De Lucia}, G., {Fontanot}, F., {Xie}, L., \& {Hirschmann}, M. 2024, \aap, 687, A68

\bibitem[{{Dekel} {et~al.}(2009{\natexlab{a}}){Dekel}, {Birnboim}, {Engel}, {Freundlich}, {Goerdt}, {Mumcuoglu}, {Neistein}, {Pichon}, {Teyssier}, \& {Zinger}}]{DekelBirnboim2009}
{Dekel}, A., {Birnboim}, Y., {Engel}, G., {et~al.} 2009{\natexlab{a}}, \nat, 457, 451

\bibitem[{{Dekel} {et~al.}(2009{\natexlab{b}}){Dekel}, {Sari}, \& {Ceverino}}]{Dekel+2009}
{Dekel}, A., {Sari}, R., \& {Ceverino}, D. 2009{\natexlab{b}}, \apj, 703, 785

\bibitem[{{Donnari} {et~al.}(2021){Donnari}, {Pillepich}, {Joshi}, {Nelson}, {Genel}, {Marinacci}, {Rodriguez-Gomez}, {Pakmor}, {Torrey}, {Vogelsberger}, \& {Hernquist}}]{Donnari+2021}
{Donnari}, M., {Pillepich}, A., {Joshi}, G.~D., {et~al.} 2021, \mnras, 500, 4004

\bibitem[{{Dubois} {et~al.}(2014){Dubois}, {Pichon}, {Welker}, {Le Borgne}, {Devriendt}, {Laigle}, {Codis}, {Pogosyan}, {Arnouts}, {Benabed}, {Bertin}, {Blaizot}, {Bouchet}, {Cardoso}, {Colombi}, {de Lapparent}, {Desjacques}, {Gavazzi}, {Kassin}, {Kimm}, {McCracken}, {Milliard}, {Peirani}, {Prunet}, {Rouberol}, {Silk}, {Slyz}, {Sousbie}, {Teyssier}, {Tresse}, {Treyer}, {Vibert}, \& {Volonteri}}]{Dubois+2014}
{Dubois}, Y., {Pichon}, C., {Welker}, C., {et~al.} 2014, \mnras, 444, 1453

\bibitem[{{Gal{\'a}rraga-Espinosa} {et~al.}(2024){Gal{\'a}rraga-Espinosa}, {Cadiou}, {Gouin}, {White}, {Springel}, {Pakmor}, {Hadzhiyska}, {Bose}, {Ferlito}, {Hernquist}, {Kannan}, {Barrera}, {Maria Delgado}, \& {Hern{\'a}ndez-Aguayo}}]{galarraga-espinosa+2024}
{Gal{\'a}rraga-Espinosa}, D., {Cadiou}, C., {Gouin}, C., {et~al.} 2024, \aap, 684, A63

\bibitem[{{Gal{\'a}rraga-Espinosa} {et~al.}(2026){Gal{\'a}rraga-Espinosa}, {Kauffmann}, {Bonoli}, {Lucie-Smith}, {Gonz{\'a}lez Delgado}, {Tempel}, {Abramo}, {Gurung-L{\'o}pez}, {Marra}, {Alcaniz}, {Benitez}, {Carneiro}, {Cenarro}, {Crist{\'o}bal-Hornillos}, {Dupke}, {Ederoclite}, {Hern{\'a}n-Caballero}, {Hern{\'a}ndez-Monteagudo}, {L{\'o}pez-Sanjuan}, {Mar{\'\i}n-Franch}, {Mendes de Oliveira}, {Moles}, {Sodr{\'e}}, {Taylor}, {Varela}, \& {V{\'a}zquez Rami{\'o}}}]{Galarraga-Espinosa+2026}
{Gal{\'a}rraga-Espinosa}, D., {Kauffmann}, G., {Bonoli}, S., {et~al.} 2026, \aap, 706, A21

\bibitem[{{Ganeshaiah Veena} {et~al.}(2018){Ganeshaiah Veena}, {Cautun}, {van de Weygaert}, {Tempel}, {Jones}, {Rieder}, \& {Frenk}}]{Ganeshaiah_Veena+2019}
{Ganeshaiah Veena}, P., {Cautun}, M., {van de Weygaert}, R., {et~al.} 2018, \mnras, 481, 414

\bibitem[{{Hahn} {et~al.}(2007){Hahn}, {Carollo}, {Porciani}, \& {Dekel}}]{Hahn+2007}
{Hahn}, O., {Carollo}, C.~M., {Porciani}, C., \& {Dekel}, A. 2007, \mnras, 381, 41

\bibitem[{{Hasan} {et~al.}(2023){Hasan}, {Burchett}, {Abeyta}, {Hellinger}, {Mandelker}, {Primack}, {Faber}, {Koo}, {Elek}, \& {Nagai}}]{Hasan+2023}
{Hasan}, F., {Burchett}, J.~N., {Abeyta}, A., {et~al.} 2023, \apj, 950, 114

\bibitem[{{Hoosain} {et~al.}(2024){Hoosain}, {Blyth}, {Skelton}, {Kannappan}, {Stark}, {Eckert}, {Hutchens}, {Carr}, \& {Kraljic}}]{Hoosain+2024}
{Hoosain}, M., {Blyth}, S.-L., {Skelton}, R.~E., {et~al.} 2024, \mnras, 528, 4139

\bibitem[{{Jego} {et~al.}(2026){Jego}, {Kraljic}, {B{\'e}thermin}, \& {Dav{\'e}}}]{Jego+2026}
{Jego}, B., {Kraljic}, K., {B{\'e}thermin}, M., \& {Dav{\'e}}, R. 2026, \aap, 706, A150

\bibitem[{{Katz} {et~al.}(2003){Katz}, {Keres}, {Dave}, \& {Weinberg}}]{Katz+2003}
{Katz}, N., {Keres}, D., {Dave}, R., \& {Weinberg}, D.~H. 2003, in Astrophysics and Space Science Library, Vol. 281, The IGM/Galaxy Connection. The Distribution of Baryons at z=0, ed. J.~L. {Rosenberg} \& M.~E. {Putman}, 185

\bibitem[{{Kere{\v{s}}} {et~al.}(2005){Kere{\v{s}}}, {Katz}, {Weinberg}, \& {Dav{\'e}}}]{Keres+2005}
{Kere{\v{s}}}, D., {Katz}, N., {Weinberg}, D.~H., \& {Dav{\'e}}, R. 2005, \mnras, 363, 2

\bibitem[{{Kleiner} {et~al.}(2017){Kleiner}, {Pimbblet}, {Jones}, {Koribalski}, \& {Serra}}]{Kleiner+2017}
{Kleiner}, D., {Pimbblet}, K.~A., {Jones}, D.~H., {Koribalski}, B.~S., \& {Serra}, P. 2017, \mnras, 466, 4692

\bibitem[{Kraljic {et~al.}(2018)Kraljic, Arnouts, Pichon, Laigle, de~la Torre, Vibert, Cadiou, Dubois, Treyer, Schimd, Codis, de~Lapparent, Devriendt, Hwang, Le~Borgne, Malavasi, Milliard, Musso, Pogosyan, Alpaslan, Bland-Hawthorn, \& Wright}]{Kraljic+2018}
Kraljic, K., Arnouts, S., Pichon, C., {et~al.} 2018, Monthly Notices of the Royal Astronomical Society, 474, 547

\bibitem[{{Kraljic} {et~al.}(2021){Kraljic}, {Duckworth}, {Tojeiro}, {Alam}, {Bizyaev}, {Weijmans}, {Boardman}, \& {Lane}}]{Kraljic+2021}
{Kraljic}, K., {Duckworth}, C., {Tojeiro}, R., {et~al.} 2021, \mnras, 504, 4626

\bibitem[{{Kuutma} {et~al.}(2017){Kuutma}, {Tamm}, \& {Tempel}}]{kuutma+2017}
{Kuutma}, T., {Tamm}, A., \& {Tempel}, E. 2017, \aap, 600, L6

\bibitem[{{Laigle} {et~al.}(2018){Laigle}, {Pichon}, {Arnouts}, {McCracken}, {Dubois}, {Devriendt}, {Slyz}, {Le Borgne}, {Benoit-L{\'e}vy}, {Hwang}, {Ilbert}, {Kraljic}, {Malavasi}, {Park}, \& {Vibert}}]{Laigle+2018}
{Laigle}, C., {Pichon}, C., {Arnouts}, S., {et~al.} 2018, \mnras, 474, 5437

\bibitem[{{Lee} {et~al.}(2017){Lee}, {Primack}, {Behroozi}, {Rodr{\'\i}guez-Puebla}, {Hellinger}, \& {Dekel}}]{Lee+2017}
{Lee}, C.~T., {Primack}, J.~R., {Behroozi}, P., {et~al.} 2017, \mnras, 466, 3834

\bibitem[{{Lemson} \& {Kauffmann}(1999)}]{Lemson+1999}
{Lemson}, G. \& {Kauffmann}, G. 1999, \mnras, 302, 111

\bibitem[{{Luber} {et~al.}(2025){Luber}, {Stierwalt}, {Privon}, {Besla}, {Johnson}, {Kallivayalil}, {Patton}, {Putman}, \& {Zhu}}]{Luber+2025}
{Luber}, N., {Stierwalt}, S., {Privon}, G.~C., {et~al.} 2025, \apjl, 993, L14

\bibitem[{{Luber} {et~al.}(2019){Luber}, {van Gorkom}, {Hess}, {Pisano}, {Fern{\'a}ndez}, \& {Momjian}}]{luber+2019}
{Luber}, N., {van Gorkom}, J.~H., {Hess}, K.~M., {et~al.} 2019, \aj, 157, 254

\bibitem[{{Malavasi} {et~al.}(2017){Malavasi}, {Arnouts}, {Vibert}, {de la Torre}, {Moutard}, {Pichon}, {Davidzon}, {Kraljic}, {Bolzonella}, {Guzzo}, {Garilli}, {Scodeggio}, {Granett}, {Abbas}, {Adami}, {Bottini}, {Cappi}, {Cucciati}, {Franzetti}, {Fritz}, {Iovino}, {Krywult}, {Le Brun}, {Le F{\`e}vre}, {Maccagni}, {Ma{\l}ek}, {Marulli}, {Polletta}, {Pollo}, {Tasca}, {Tojeiro}, {Vergani}, {Zanichelli}, {Bel}, {Branchini}, {Coupon}, {De Lucia}, {Dubois}, {Hawken}, {Ilbert}, {Laigle}, {Moscardini}, {Sousbie}, {Treyer}, \& {Zamorani}}]{Malavasi+2017}
{Malavasi}, N., {Arnouts}, S., {Vibert}, D., {et~al.} 2017, \mnras, 465, 3817

\bibitem[{{Marinacci} {et~al.}(2018){Marinacci}, {Vogelsberger}, {Pakmor}, {Torrey}, {Springel}, {Hernquist}, {Nelson}, {Weinberger}, {Pillepich}, {Naiman}, \& {Genel}}]{Illustris5}
{Marinacci}, F., {Vogelsberger}, M., {Pakmor}, R., {et~al.} 2018, \mnras, 480, 5113

\bibitem[{{Martizzi} {et~al.}(2019){Martizzi}, {Vogelsberger}, {Artale}, {Haider}, {Torrey}, {Marinacci}, {Nelson}, {Pillepich}, {Weinberger}, {Hernquist}, {Naiman}, \& {Springel}}]{Martizzi+2019}
{Martizzi}, D., {Vogelsberger}, M., {Artale}, M.~C., {et~al.} 2019, \mnras, 486, 3766

\bibitem[{{Mo} {et~al.}(1998){Mo}, {Mao}, \& {White}}]{Mo+1998}
{Mo}, H.~J., {Mao}, S., \& {White}, S. D.~M. 1998, \mnras, 295, 319

\bibitem[{{Mo} \& {White}(1996)}]{Mo+1996}
{Mo}, H.~J. \& {White}, S.~D.~M. 1996, \mnras, 282, 347

\bibitem[{{Muratov} {et~al.}(2015){Muratov}, {Kere{\v{s}}}, {Faucher-Gigu{\`e}re}, {Hopkins}, {Quataert}, \& {Murray}}]{Muratov+2015}
{Muratov}, A.~L., {Kere{\v{s}}}, D., {Faucher-Gigu{\`e}re}, C.-A., {et~al.} 2015, \mnras, 454, 2691

\bibitem[{{Musso} {et~al.}(2018){Musso}, {Cadiou}, {Pichon}, {Codis}, {Kraljic}, \& {Dubois}}]{Musso+2018}
{Musso}, M., {Cadiou}, C., {Pichon}, C., {et~al.} 2018, \mnras, 476, 4877

\bibitem[{{Naiman} {et~al.}(2018){Naiman}, {Pillepich}, {Springel}, {Ramirez-Ruiz}, {Torrey}, {Vogelsberger}, {Pakmor}, {Nelson}, {Marinacci}, {Hernquist}, {Weinberger}, \& {Genel}}]{Illustris4}
{Naiman}, J.~P., {Pillepich}, A., {Springel}, V., {et~al.} 2018, \mnras, 477, 1206

\bibitem[{{Navdha} {et~al.}(2025){Navdha}, {Busch}, \& {White}}]{Navdha+2025}
{Navdha}, {Busch}, P., \& {White}, S. D.~M. 2025, \mnras, 539, 1248

\bibitem[{{Nelson} {et~al.}(2018){Nelson}, {Pillepich}, {Springel}, {Weinberger}, {Hernquist}, {Pakmor}, {Genel}, {Torrey}, {Vogelsberger}, {Kauffmann}, {Marinacci}, \& {Naiman}}]{Illustris1}
{Nelson}, D., {Pillepich}, A., {Springel}, V., {et~al.} 2018, \mnras, 475, 624

\bibitem[{{O'Kane} {et~al.}(2024){O'Kane}, {Kuchner}, {Gray}, \& {Arag{\'o}n-Salamanca}}]{OKane+2024}
{O'Kane}, C.~J., {Kuchner}, U., {Gray}, M.~E., \& {Arag{\'o}n-Salamanca}, A. 2024, \mnras, 534, 1682

\bibitem[{{Pasha} {et~al.}(2023){Pasha}, {Mandelker}, {van den Bosch}, {Springel}, \& {van de Voort}}]{Pasha+2023}
{Pasha}, I., {Mandelker}, N., {van den Bosch}, F.~C., {Springel}, V., \& {van de Voort}, F. 2023, \mnras, 520, 2692

\bibitem[{{Pichon} {et~al.}(2011){Pichon}, {Pogosyan}, {Kimm}, {Slyz}, {Devriendt}, \& {Dubois}}]{Pichon+2011}
{Pichon}, C., {Pogosyan}, D., {Kimm}, T., {et~al.} 2011, \mnras, 418, 2493

\bibitem[{{Pillepich} {et~al.}(2018{\natexlab{a}}){Pillepich}, {Nelson}, {Hernquist}, {Springel}, {Pakmor}, {Torrey}, {Weinberger}, {Genel}, {Naiman}, {Marinacci}, \& {Vogelsberger}}]{Illustris3}
{Pillepich}, A., {Nelson}, D., {Hernquist}, L., {et~al.} 2018{\natexlab{a}}, \mnras, 475, 648

\bibitem[{{Pillepich} {et~al.}(2019){Pillepich}, {Nelson}, {Springel}, {Pakmor}, {Torrey}, {Weinberger}, {Vogelsberger}, {Marinacci}, {Genel}, {van der Wel}, \& {Hernquist}}]{Pillepich+2019_sfr}
{Pillepich}, A., {Nelson}, D., {Springel}, V., {et~al.} 2019, \mnras, 490, 3196

\bibitem[{{Pillepich} {et~al.}(2018{\natexlab{b}}){Pillepich}, {Springel}, {Nelson}, {Genel}, {Naiman}, {Pakmor}, {Hernquist}, {Torrey}, {Vogelsberger}, {Weinberger}, \& {Marinacci}}]{Pillepich+2018}
{Pillepich}, A., {Springel}, V., {Nelson}, D., {et~al.} 2018{\natexlab{b}}, \mnras, 473, 4077

\bibitem[{{Planck Collaboration} {et~al.}(2016){Planck Collaboration}, {Ade}, {Aghanim}, {Arnaud}, {Ashdown}, {Aumont}, {Baccigalupi}, {Banday}, {Barreiro}, {Bartlett}, {Bartolo}, {Battaner}, {Battye}, {Benabed}, {Beno{\^\i}t}, {Benoit-L{\'e}vy}, {Bernard}, {Bersanelli}, {Bielewicz}, {Bock}, {Bonaldi}, {Bonavera}, {Bond}, {Borrill}, {Bouchet}, {Boulanger}, {Bucher}, {Burigana}, {Butler}, {Calabrese}, {Cardoso}, {Catalano}, {Challinor}, {Chamballu}, {Chary}, {Chiang}, {Chluba}, {Christensen}, {Church}, {Clements}, {Colombi}, {Colombo}, {Combet}, {Coulais}, {Crill}, {Curto}, {Cuttaia}, {Danese}, {Davies}, {Davis}, {de Bernardis}, {de Rosa}, {de Zotti}, {Delabrouille}, {D{\'e}sert}, {Di Valentino}, {Dickinson}, {Diego}, {Dolag}, {Dole}, {Donzelli}, {Dor{\'e}}, {Douspis}, {Ducout}, {Dunkley}, {Dupac}, {Efstathiou}, {Elsner}, {En{\ss}lin}, {Eriksen}, {Farhang}, {Fergusson}, {Finelli}, {Forni}, {Frailis}, {Fraisse}, {Franceschi}, {Frejsel}, {Galeotta}, {Galli}, {Ganga}, {Gauthier}, {Gerbino}, {Ghosh}, {Giard},
  {Giraud-H{\'e}raud}, {Giusarma}, {Gjerl{\o}w}, {Gonz{\'a}lez-Nuevo}, {G{\'o}rski}, {Gratton}, {Gregorio}, {Gruppuso}, {Gudmundsson}, {Hamann}, {Hansen}, {Hanson}, {Harrison}, {Helou}, {Henrot-Versill{\'e}}, {Hern{\'a}ndez-Monteagudo}, {Herranz}, {Hildebrandt}, {Hivon}, {Hobson}, {Holmes}, {Hornstrup}, {Hovest}, {Huang}, {Huffenberger}, {Hurier}, {Jaffe}, {Jaffe}, {Jones}, {Juvela}, {Keih{\"a}nen}, {Keskitalo}, {Kisner}, {Kneissl}, {Knoche}, {Knox}, {Kunz}, {Kurki-Suonio}, {Lagache}, {L{\"a}hteenm{\"a}ki}, {Lamarre}, {Lasenby}, {Lattanzi}, {Lawrence}, {Leahy}, {Leonardi}, {Lesgourgues}, {Levrier}, {Lewis}, {Liguori}, {Lilje}, {Linden-V{\o}rnle}, {L{\'o}pez-Caniego}, {Lubin}, {Mac{\'\i}as-P{\'e}rez}, {Maggio}, {Maino}, {Mandolesi}, {Mangilli}, {Marchini}, {Maris}, {Martin}, {Martinelli}, {Mart{\'\i}nez-Gonz{\'a}lez}, {Masi}, {Matarrese}, {McGehee}, {Meinhold}, {Melchiorri}, {Melin}, {Mendes}, {Mennella}, {Migliaccio}, {Millea}, {Mitra}, {Miville-Desch{\^e}nes}, {Moneti}, {Montier}, {Morgante}, {Mortlock},
  {Moss}, {Munshi}, {Murphy}, {Naselsky}, {Nati}, {Natoli}, {Netterfield}, {N{\o}rgaard-Nielsen}, {Noviello}, {Novikov}, {Novikov}, {Oxborrow}, {Paci}, {Pagano}, {Pajot}, {Paladini}, {Paoletti}, {Partridge}, {Pasian}, {Patanchon}, {Pearson}, {Perdereau}, {Perotto}, {Perrotta}, {Pettorino}, {Piacentini}, {Piat}, {Pierpaoli}, {Pietrobon}, {Plaszczynski}, {Pointecouteau}, {Polenta}, {Popa}, {Pratt}, {Pr{\'e}zeau}, {Prunet}, {Puget}, {Rachen}, {Reach}, {Rebolo}, {Reinecke}, {Remazeilles}, {Renault}, {Renzi}, {Ristorcelli}, {Rocha}, {Rosset}, {Rossetti}, {Roudier}, {Rouill{\'e} d'Orfeuil}, {Rowan-Robinson}, {Rubi{\~n}o-Mart{\'\i}n}, {Rusholme}, {Said}, {Salvatelli}, {Salvati}, {Sandri}, {Santos}, {Savelainen}, {Savini}, {Scott}, {Seiffert}, {Serra}, {Shellard}, {Spencer}, {Spinelli}, {Stolyarov}, {Stompor}, {Sudiwala}, {Sunyaev}, {Sutton}, {Suur-Uski}, {Sygnet}, {Tauber}, {Terenzi}, {Toffolatti}, {Tomasi}, {Tristram}, {Trombetti}, {Tucci}, {Tuovinen}, {T{\"u}rler}, {Umana}, {Valenziano}, {Valiviita}, {Van Tent},
  {Vielva}, {Villa}, {Wade}, {Wandelt}, {Wehus}, {White}, {White}, {Wilkinson}, {Yvon}, {Zacchei}, \& {Zonca}}]{Planck_2016}
{Planck Collaboration}, {Ade}, P.~A.~R., {Aghanim}, N., {et~al.} 2016, \aap, 594, A13

\bibitem[{{Proshina} \& {Oparin}(2025)}]{Proshina+2025}
{Proshina}, I.~S. \& {Oparin}, D.~V. 2025, \apj, 980, 118

\bibitem[{{Reshetnikov} \& {Combes}(1998)}]{Reshetnikov+1999}
{Reshetnikov}, V. \& {Combes}, F. 1998, \aap, 337, 9

\bibitem[{{Roberts} {et~al.}(2023){Roberts}, {Brown}, {Zabel}, {Wilson}, {Chung}, {Parker}, {Bisaria}, {Boselli}, {Catinella}, {Chown}, {Cortese}, {Davis}, {Ellison}, {Jim{\'e}nez-Donaire}, {Lee}, {Smith}, {Spekkens}, {Stevens}, {Thorp}, {Villanueva}, {Watts}, {Welker}, \& {Yoon}}]{Roberts+2023}
{Roberts}, I.~D., {Brown}, T., {Zabel}, N., {et~al.} 2023, \aap, 675, A78

\bibitem[{{Roediger} \& {Br{\"u}ggen}(2006)}]{Roediger+2006}
{Roediger}, E. \& {Br{\"u}ggen}, M. 2006, \mnras, 369, 567

\bibitem[{{Sarron} {et~al.}(2019){Sarron}, {Adami}, {Durret}, \& {Laigle}}]{Sarron+2019}
{Sarron}, F., {Adami}, C., {Durret}, F., \& {Laigle}, C. 2019, \aap, 632, A49

\bibitem[{{Sousbie}(2011)}]{Sousbie+2011}
{Sousbie}, T. 2011, \mnras, 414, 350

\bibitem[{{Sousbie} {et~al.}(2011){Sousbie}, {Pichon}, \& {Kawahara}}]{Sousbie_etal+2011}
{Sousbie}, T., {Pichon}, C., \& {Kawahara}, H. 2011, \mnras, 414, 384

\bibitem[{{Springel} {et~al.}(2018){Springel}, {Pakmor}, {Pillepich}, {Weinberger}, {Nelson}, {Hernquist}, {Vogelsberger}, {Genel}, {Torrey}, {Marinacci}, \& {Naiman}}]{Illustris2}
{Springel}, V., {Pakmor}, R., {Pillepich}, A., {et~al.} 2018, \mnras, 475, 676

\bibitem[{{Steinhauser} {et~al.}(2016){Steinhauser}, {Schindler}, \& {Springel}}]{Steinhauser+2016}
{Steinhauser}, D., {Schindler}, S., \& {Springel}, V. 2016, \aap, 591, A51

\bibitem[{{Tempel} {et~al.}(2014{\natexlab{a}}){Tempel}, {Kipper}, {Saar}, {Bussov}, {Hektor}, \& {Pelt}}]{Tempel+2014_perls}
{Tempel}, E., {Kipper}, R., {Saar}, E., {et~al.} 2014{\natexlab{a}}, \aap, 572, A8

\bibitem[{{Tempel} {et~al.}(2014{\natexlab{b}}){Tempel}, {Stoica}, {Mart{\'\i}nez}, {Liivam{\"a}gi}, {Castellan}, \& {Saar}}]{Tempel+2014}
{Tempel}, E., {Stoica}, R.~S., {Mart{\'\i}nez}, V.~J., {et~al.} 2014{\natexlab{b}}, \mnras, 438, 3465

\bibitem[{{Vera-Ciro} {et~al.}(2011){Vera-Ciro}, {Sales}, {Helmi}, {Frenk}, {Navarro}, {Springel}, {Vogelsberger}, \& {White}}]{Vera-Ciro+2011}
{Vera-Ciro}, C.~A., {Sales}, L.~V., {Helmi}, A., {et~al.} 2011, \mnras, 416, 1377

\bibitem[{{Vulcani} {et~al.}(2026){Vulcani}, {De Lucia}, {Zakharova}, {Serra}, {Xie}, {Barsanti}, {Maria Poggianti}, {Moretti}, {Gullieuszik}, {Bah{\'e}}, {Fontanot}, {Fritz}, {Gastaldello}, {Gaspari}, {Hirschmann}, {Jaff{\'e}}, {Kolokythas}, {Ignesti}, {Lassen}, {Loni}, {Lovisari}, {Marasco}, {Makhathini}, {McGee}, {Mogotsi}, {Pisano}, {Ramatsoku}, {Smirnov}, {Smith}, {Tonnesen}, \& {Verheijen}}]{Vulcani_MAGNET+2026}
{Vulcani}, B., {De Lucia}, G., {Zakharova}, D., {et~al.} 2026, \aap, 708, A141

\bibitem[{{Vulcani} {et~al.}(2021){Vulcani}, {Poggianti}, {Moretti}, {Franchetto}, {Bacchini}, {McGee}, {Jaff{\'e}}, {Mingozzi}, {Werle}, {Tomi{\v{c}}i{\'c}}, {Fritz}, {Bettoni}, {Wolter}, \& {Gullieuszik}}]{Vulcani+2021}
{Vulcani}, B., {Poggianti}, B.~M., {Moretti}, A., {et~al.} 2021, \apj, 914, 27

\bibitem[{{Vulcani} {et~al.}(2019){Vulcani}, {Poggianti}, {Moretti}, {Gullieuszik}, {Fritz}, {Franchetto}, {Fasano}, {Bettoni}, \& {Jaff{\'e}}}]{Vulcani+2019}
{Vulcani}, B., {Poggianti}, B.~M., {Moretti}, A., {et~al.} 2019, \mnras, 487, 2278

\bibitem[{{Vulcani} {et~al.}(2018){Vulcani}, {Poggianti}, {Moretti}, {Mapelli}, {Fasano}, {Fritz}, {Jaff{\'e}}, {Bettoni}, {Gullieuszik}, \& {Bellhouse}}]{Vulcani+2018_Lopsided}
{Vulcani}, B., {Poggianti}, B.~M., {Moretti}, A., {et~al.} 2018, \apj, 852, 94

\bibitem[{{Wang} {et~al.}(2011){Wang}, {Mo}, {Jing}, {Yang}, \& {Wang}}]{Wang+2011}
{Wang}, H., {Mo}, H.~J., {Jing}, Y.~P., {Yang}, X., \& {Wang}, Y. 2011, \mnras, 413, 1973

\bibitem[{{Welker} {et~al.}(2014){Welker}, {Devriendt}, {Dubois}, {Pichon}, \& {Peirani}}]{Welker+2014}
{Welker}, C., {Devriendt}, J., {Dubois}, Y., {Pichon}, C., \& {Peirani}, S. 2014, \mnras, 445, L46

\bibitem[{{Winkel} {et~al.}(2021){Winkel}, {Pasquali}, {Kraljic}, {Smith}, {Gallazzi}, \& {Jackson}}]{Winkel+2021}
{Winkel}, N., {Pasquali}, A., {Kraljic}, K., {et~al.} 2021, \mnras, 505, 4920

\bibitem[{{Wu} {et~al.}(2026){Wu}, {Wang}, \& {Liao}}]{Wu+2026}
{Wu}, Z., {Wang}, Y., \& {Liao}, S. 2026, Research in Astronomy and Astrophysics, 26, 095019

\bibitem[{{Xie} {et~al.}(2020){Xie}, {De Lucia}, {Hirschmann}, \& {Fontanot}}]{Xie+2020}
{Xie}, L., {De Lucia}, G., {Hirschmann}, M., \& {Fontanot}, F. 2020, \mnras, 498, 4327

\bibitem[{{Yu} {et~al.}(2025){Yu}, {Zhu}, {Yang}, {Mo}, {Luan}, \& {Feng}}]{Yu+2025}
{Yu}, G., {Zhu}, W., {Yang}, Q.-R., {et~al.} 2025, \apj, 986, 193

\bibitem[{{Zakharova} {et~al.}(2026){Zakharova}, {De Lucia}, {Vulcani}, {Fontanot}, \& {Xie}}]{Zakharova_envI}
{Zakharova}, D., {De Lucia}, G., {Vulcani}, B., {Fontanot}, F., \& {Xie}, L. 2026, \aap, 707, A377

\bibitem[{{Zakharova} {et~al.}(2024){Zakharova}, {Vulcani}, {De Lucia}, {Finn}, {Rudnick}, {Combes}, {Castignani}, {Fontanot}, {Jablonka}, {Xie}, \& {Hirschmann}}]{Zakharova+2024}
{Zakharova}, D., {Vulcani}, B., {De Lucia}, G., {et~al.} 2024, \aap, 690, A300

\bibitem[{{Zakharova} {et~al.}(2023){Zakharova}, {Vulcani}, {De Lucia}, {Xie}, {Hirschmann}, \& {Fontanot}}]{Zakharova+2023}
{Zakharova}, D., {Vulcani}, B., {De Lucia}, G., {et~al.} 2023, \mnras, 525, 4079

\bibitem[{{Zarattini} \& {Aguerri}(2025)}]{Zarattini+2025}
{Zarattini}, S. \& {Aguerri}, J.~A.~L. 2025, \aap, 698, A196

\bibitem[{{Zaritsky} {et~al.}(2013){Zaritsky}, {Salo}, {Laurikainen}, {Elmegreen}, {Athanassoula}, {Bosma}, {Comer{\'o}n}, {Erroz-Ferrer}, {Elmegreen}, {Gadotti}, {Gil de Paz}, {Hinz}, {Ho}, {Holwerda}, {Kim}, {Knapen}, {Laine}, {Laine}, {Madore}, {Meidt}, {Menendez-Delmestre}, {Mizusawa}, {Mu{\~n}oz-Mateos}, {Regan}, {Seibert}, \& {Sheth}}]{Zaritsky+2013}
{Zaritsky}, D., {Salo}, H., {Laurikainen}, E., {et~al.} 2013, \apj, 772, 135

\bibitem[{{Zhu} {et~al.}(2021){Zhu}, {Zhang}, \& {Feng}}]{Zhu+2021}
{Zhu}, W., {Zhang}, F., \& {Feng}, L.-L. 2021, \apj, 920, 2

\end{thebibliography}

\begin{appendix}

\section{Stellar and halo mass assembly}
\label{app:stm_sf_interg}

To avoid repeating the detailed analysis presented in \citet{Zakharova_envI}, we briefly verify here that their results are reproduced in TNG50 when extending the sample to lower-mass galaxies $\rm 8 \le \log(M_{star}/M_{sun}) \le 10 $.
\begin{figure}[h!]
    \centering
    \includegraphics[width=1\linewidth]{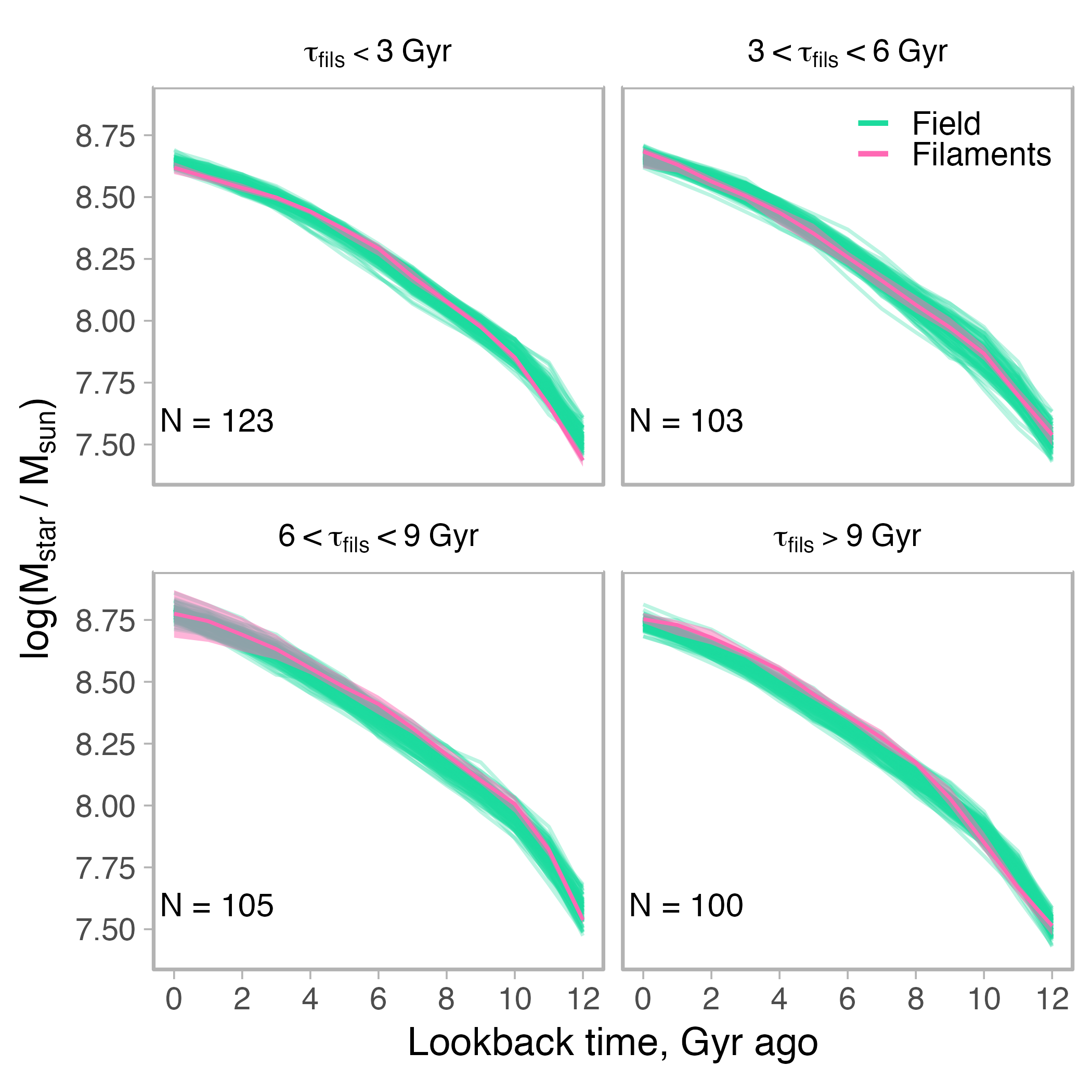}
    \caption{Stellar mass assembly histories of present-day low-mass filament galaxies as a function of infall time, compared with their stellar- and halo-mass–matched field galaxies~(100 realisations).}  
    \label{fig:stm_evolution}
\end{figure}


\begin{figure}[h!]
    \centering
    \includegraphics[width=1\linewidth]{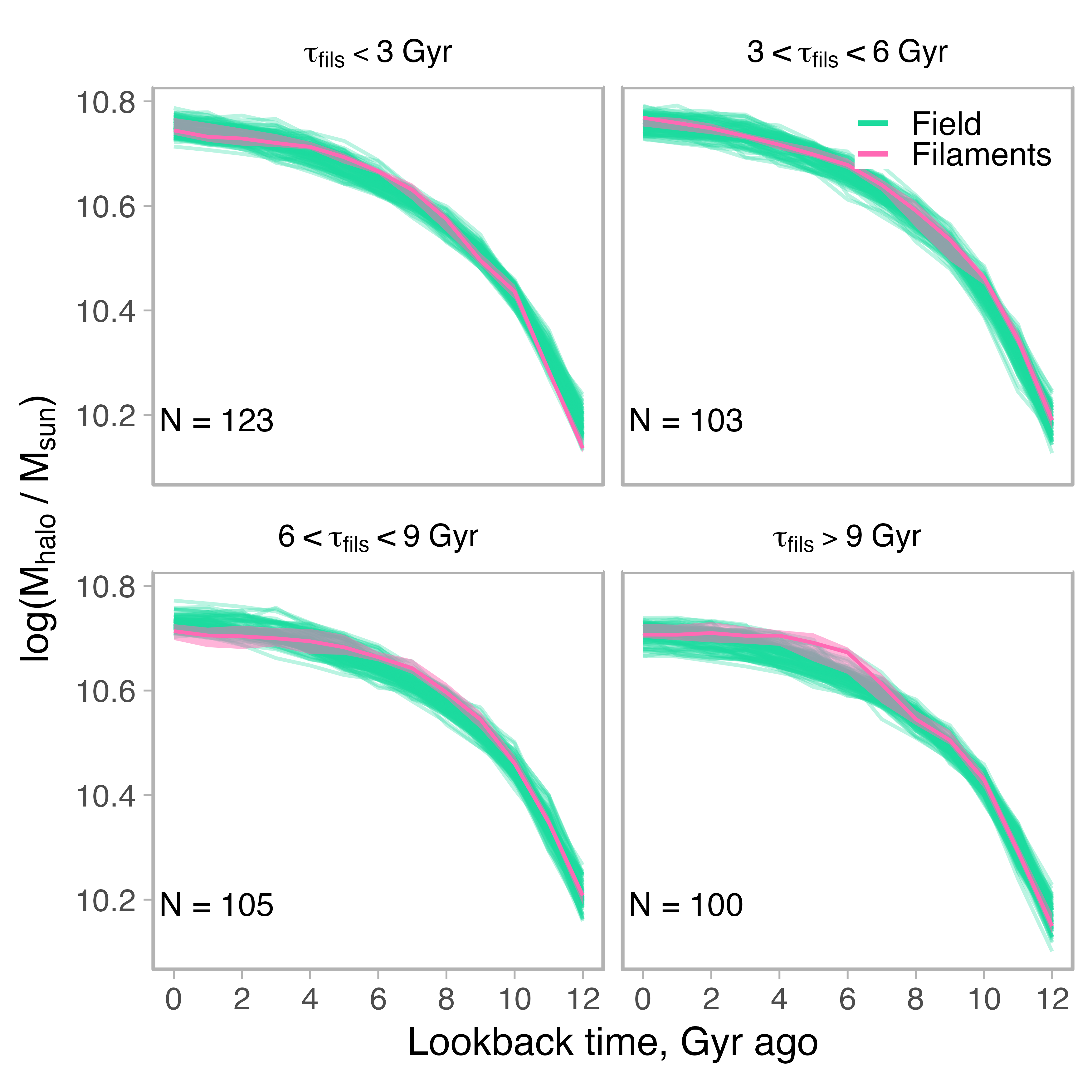}
    \caption{Same as Fig.~\ref{fig:stm_evolution} but for halo assembly.  }
    \label{fig:halo_evolution}
\end{figure}


Figures~\ref{fig:stm_evolution} and \ref{fig:halo_evolution} show the evolution of stellar and halo mass assembly for filament and field galaxies, matched in both stellar and halo mass at z=0. In all panels, filament galaxies exhibit mass growth consistent with the spread of the field population. This indicates that low-mass filament and field galaxies have similar stellar and halo mass assembly histories, regardless of infall time. This result is consistent with \citet{Zakharova_envI}, who found the same behaviour in both TNG100 and GAEA when controlling for stellar and halo mass at z=0.

\par
Therefore, TNG50-1 predicts no significant difference in the stellar or halo mass assembly of galaxies in filaments compared to those in the field, consistent with \citet{Zakharova_envI}.

\section{Mass $R_{50\%,cold~gas}$ size relations }
\label{app:mass-size50}

\begin{figure}[!ht]
    \centering
    \includegraphics[width=1\linewidth]{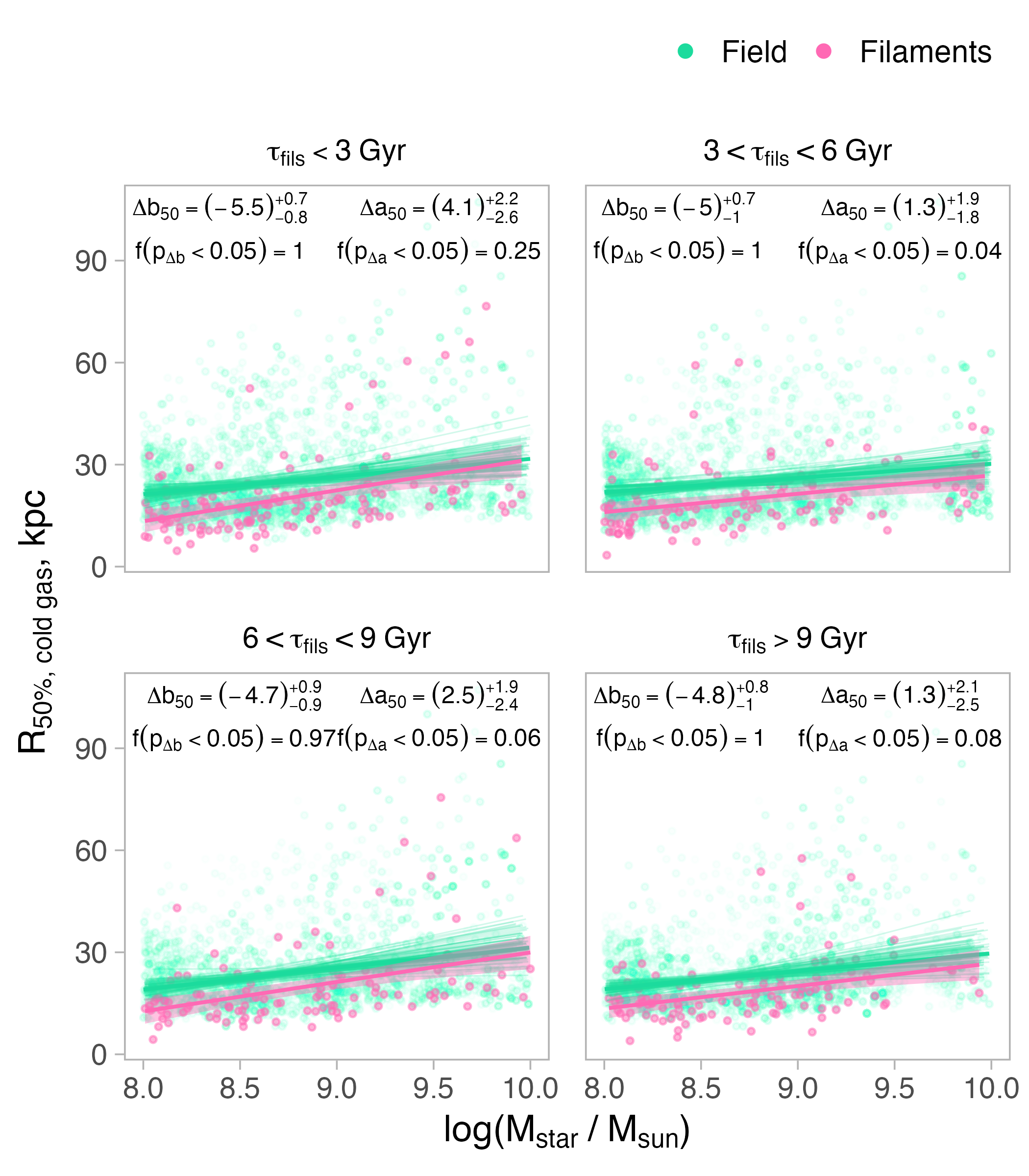}
    \caption{Same as Fig.~\ref{fig:r90_cold_gas_z0} but for $R_{50\%,cold~gas}$  }
    \label{fig:r50_cold_gas_z0}
\end{figure}

In Section~\ref{sec:mass-r90}, we found that the size of the cold gas component $R_{90\%,cold~gas}$ is smaller for galaxies inside filaments compared to the field. Here, we provide an analysis of the inner parts of galaxies and investigate, in the same manner, $R_{50\%, cold~gas}$ as a function of stellar mass at z=0. Fig.~\ref{fig:r50_cold_gas_z0} is an analogue of the original Fig.~\ref{fig:r90_cold_gas_z0}. It again shows that all filament points lie below the 100 field realisations, with support from the ANCOVA test, which demonstrates the statistical significance of the difference in intercept $\Delta b$ in 99--100\% of cases. The median difference in intercept is around 7 kpc within uncertainties for filament galaxies with different $\tau_{\mathrm{fils}}$. Therefore, we conclude that galaxies have smaller cold disc sizes not only on the periphery but also in the central 30-50 kpc.

\section{Examples of cold-gas morphology for the different cold gas evolution of low-mass galaxies}
\label{app:galaxies}

\begin{figure*}[!ht]
    \centering
    \includegraphics[width=1\linewidth]{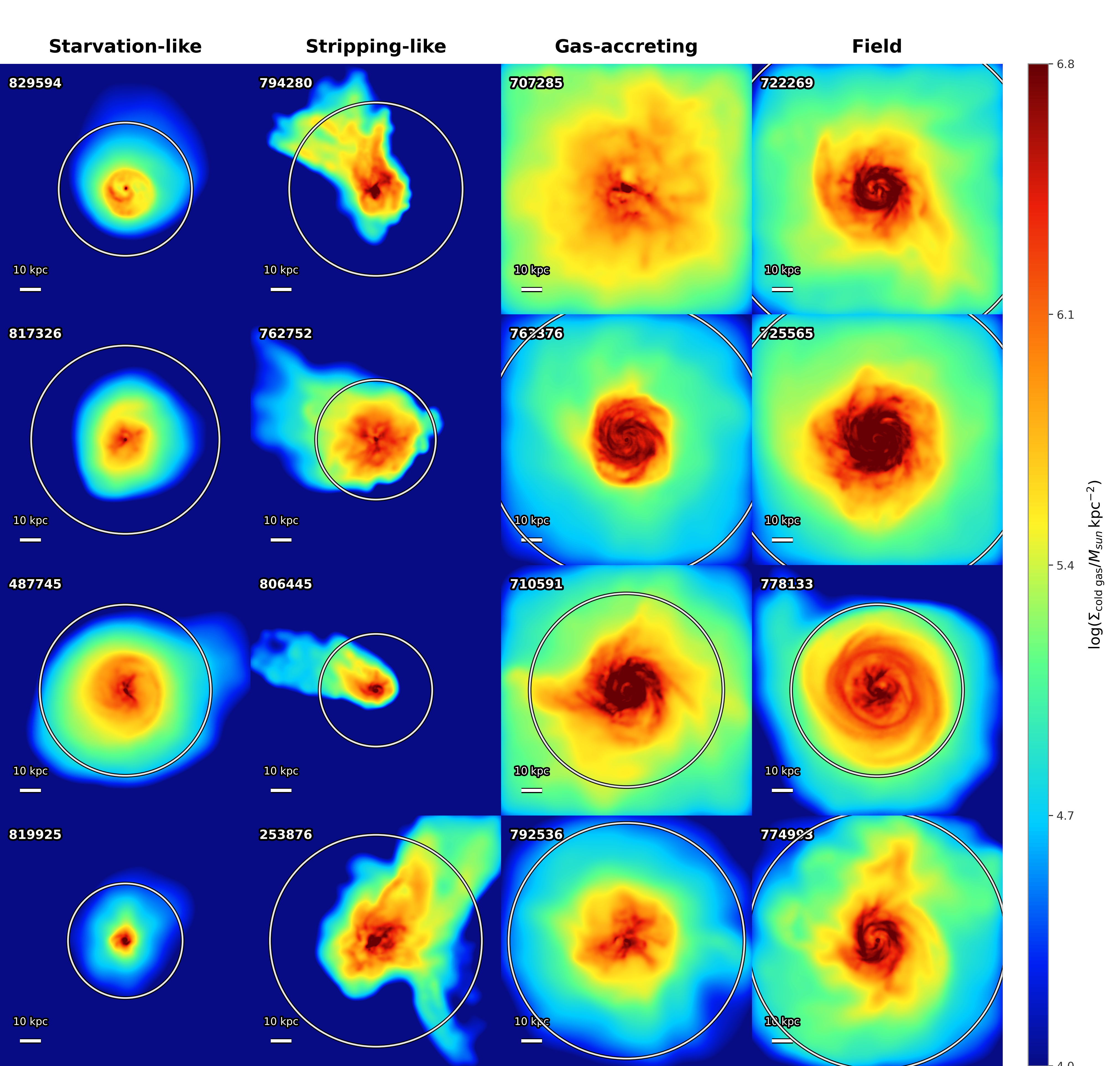}
    \caption{Examples of galaxies classified as described in Sec.~\ref{subsec:mass_classification}, compared to field galaxies at z=0. Columns show starvation-like, stripping-like, gas-accreting, and field galaxies. The field-of-view size is the same across panels for comparison. The white circle demonstrates a $R_{90\%,\mathrm{cold~gas}}$ at z=0. Labels give the subhalo IDs. }
    \label{fig:galaxies_example}
\end{figure*}


\textcolor{black}{In this Appendix, we show examples of galaxies classified using the method described in Sec.~\ref{subsec:mass_classification}. We show starvation-like, stripping-like, and gas-accreting galaxies, and compare them to galaxies in the field. This figure provides an illustration of the different cold-gas morphologies associated with each gas evolution scenario. All panels share the same field of view and colour scale, and the white circle marks $R_{90\%,\mathrm{cold~gas}}$. These are representative examples; the quantitative differences are given in Figs.~\ref{fig:r90_cold_gas_z0}, ~\ref{fig:size_evolution_example},  and ~\ref{fig:starvation_vs_stripping_example}.}
\par
\textcolor{black}{The first column represents the starved population of galaxies inside filaments. They have smaller discs than field galaxies (fourth column) and little to no internal structure. The second column shows galaxies classified as stripping-like, which possess a tail of cold gas that corresponds to fast gas removal. The third column shows galaxies within filaments that continue to accrete gas after infall; they enlarged their radii and developed internal structure, unlike the starved population. The fourth column shows field galaxies, presented for comparison.}

\section{Cold gas content of field galaxies}
\label{app:starvation_vs_stripping_field}

\begin{figure}[h!]
    \centering
    \includegraphics[width=1\linewidth]{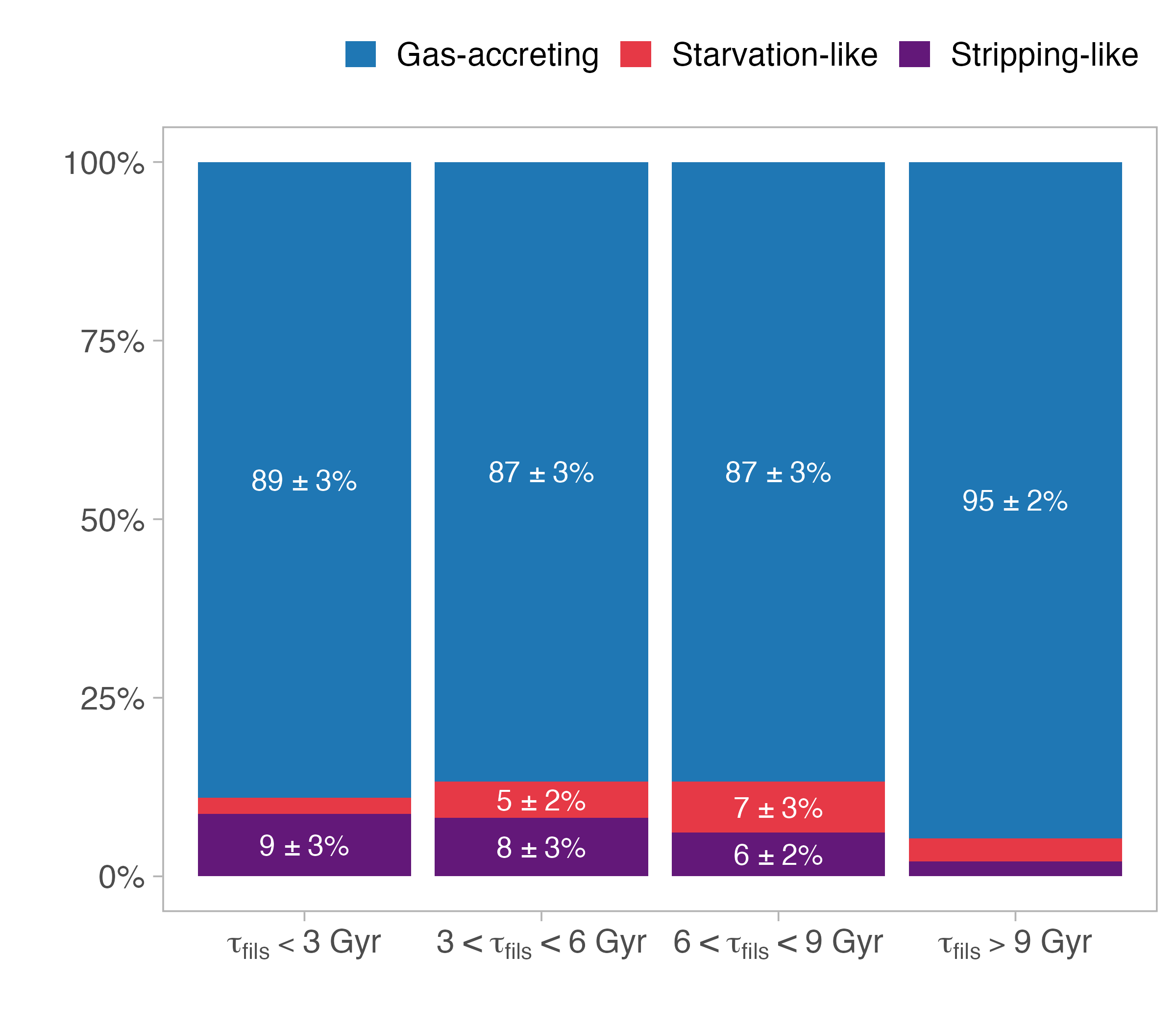}
    \caption{Same as Fig.~\ref{fig:starvation_vs_stripping_analysis}, but for field galaxies. Each bar represents 100 stellar- and halo-mass–matched realisations of the field sample, constructed to match the filament population, with mock infall times assigned to reproduce the corresponding $\tau_{\mathrm{fils}}$ bins.  }
    \label{fig:starvation_vs_stripping_analysis_field}
\end{figure}

To assess whether the trends identified for filament galaxies are driven by the filaments, we apply the same classification scheme described in Section~\ref{subsec:mass_classification} to a control sample of field galaxies. Fig.~\ref{fig:starvation_vs_stripping_analysis_field} shows that the vast majority of field galaxies, matched in stellar and halo mass, are consistent with continued gas accretion. Only $\sim$5–10\% of the field population exhibits signatures of starvation-like or stripping-like evolution. In the field, the small fraction of stripping-like cases may be associated with galaxy–galaxy interactions, which can induce gas removal during close passages. A small fraction of galaxies also shows starvation-like evolution without an obvious environmental driver. Nevertheless, the fraction of both starvation-like and stripping-like systems in the field remains significantly lower than in filaments. This demonstrates that the diversity of gas evolution scenarios identified in the main analysis is not an intrinsic feature of low-mass galaxy evolution, but is instead driven by environmental processes operating within filaments.


\section{Gas morphology}
\label{sec::assym}

The mechanisms discussed above not only affect the amount or extent of cold gas but can also leave imprints on the morphology of the gas discs. Several environmental processes have been shown to induce lopsided gas and stellar discs, such as stripping \citep[see e.g.][]{Reshetnikov+1999, Zaritsky+2013, Roberts+2023, Bilimogga+2026, Proshina+2025} and gas accretion from the cosmic web~\citep{Bournaud+2005, Vulcani+2018_Lopsided}. The asymmetry of gas discs can also be estimated in observations, allowing for a direct comparison of model predictions with observations. In this section, we examine the morphology of gas discs at z=0 for galaxies residing in filaments.
\par
We estimate the asymmetry of the cold-gas disc by projecting the 3D gas distribution onto a plane and constructing a 2D map of the cold-gas surface density, $I$. We then compare this map with the same map rotated by $180^{\circ}$, $I_{180}$. The rotational asymmetry parameter is defined as
\begin{equation}
A_{\Sigma} = \frac{\sum |I - I_{180}|}{2 \sum |I|}.
\end{equation}
We estimate this parameter within a radial ring between $R_{67\%,\mathrm{cold~gas}}$ and $R_{90\%,\mathrm{cold~gas}}$, in order to exclude the typically symmetric central regions and the highly noisy outermost gas. However, we have also tested the same analysis in the rings $0$--$R_{67\%,\mathrm{cold~gas}}$ and $R_{90\%,\mathrm{cold~gas}}$--$R_{99\%,\mathrm{cold~gas}}$, and found that the results do not affect our conclusions.

\begin{figure}
    \centering
    \includegraphics[width=1\linewidth]{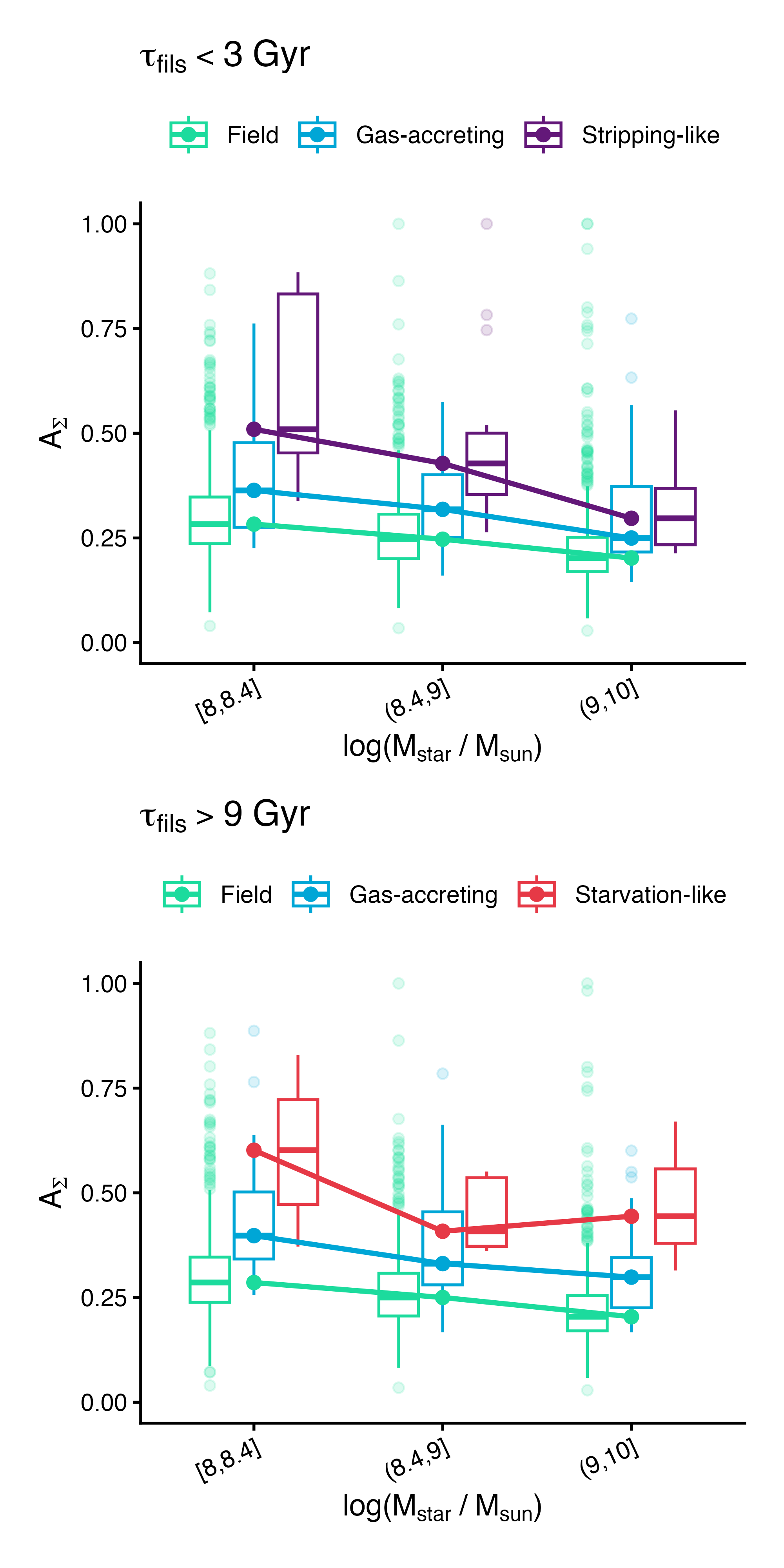}
    \caption{The rotational asymmetry parameter for field and filament galaxies  with $\tau_{\mathrm{fils}} < 3~\mathrm{Gyr}$ and $\tau_{\mathrm{fils}} > 9~\mathrm{Gyr}$. The samples are matched to have the same stellar mass and halo mass distribution at  z=0. }

    \label{fig:assymentry_as_stm}
\end{figure}


Fig.~\ref{fig:assymentry_as_stm} shows the asymmetry parameter of the cold-gas surface density, $A_{\Sigma}$, as a function of stellar mass for filament galaxies that experienced different gas evolution. The top panel shows $A_{\Sigma}$ for recent infallers into filaments ($\tau_{\mathrm{fils}} < 3~\mathrm{Gyr}$) and for a stellar and halo mass-matched sample of field galaxies. $A_{\Sigma}$ depends on stellar mass for all three galaxy populations. However, field galaxies exhibit the lowest asymmetry, while the stripped population shows the highest values, with a particularly large scatter at the low-mass end.  The high lopsidedness of the stripped population is expected, since stripping is typically more efficient on one side of the galaxy, producing strong asymmetries. In addition,  low-mass galaxies are expected to be more strongly affected by ram-pressure stripping in groups and clusters due to their shallower gravitational potentials \citep{Donnari+2021, Xie+2020}. Finally, as shown in Fig.~\ref{fig:r90_cold_gas_z0}, low-mass galaxies tend to have smaller gas discs. Therefore, the behaviour shown in Fig.~\ref{fig:assymentry_as_stm} for filament galaxies experiencing a stripping-like evolution is consistent with the cosmic web stripping.
\par
Filament galaxies $\tau_{\mathrm{fils}} < 3~\mathrm{Gyr}$ that are still accreting gas exhibit slightly higher median values of $A_{\Sigma}$ than field galaxies, although the difference remains within the uncertainties. Therefore, while the tidal field of the cosmic web may affect the gas morphology of these galaxies, the impact is relatively modest.
\par
The bottom panel of Fig.~\ref{fig:assymentry_as_stm} shows $A_{\Sigma}$ for galaxies with $\tau_{\mathrm{fils}} > 9~\mathrm{Gyr}$ and for their stellar-mass-matched field counterparts. Again, field galaxies exhibit the lowest values of $A_{\Sigma}$ across all stellar mass bins, while the starved population shows the highest asymmetry, with gas-accreting galaxies lying at intermediate values. Galaxies that experienced a starvation-like process after infall exhibit higher $A_{\Sigma}$ even than the stripped population. Although this seems counterintuitive, galaxies that entered filaments more than $9~\mathrm{Gyr}$ ago did so at a time when all galaxies displayed larger gas-disc asymmetries. After infall, these galaxies became unable to accrete new gas and became a "stalled" population, effectively preserving their more prominent asymmetry. In contrast, both gas-accreting filament galaxies and field galaxies continue accreting gas. This ongoing inflow can replenish the gas disc and promote azimuthal redistribution of the cold gas, leading to a smoother and more regular gas distribution. However, as we show in Sec.~\ref{subsec:mass_classification}, the gas-accreting population does not redistribute its gas efficiently to larger radii. It is therefore possible that the same mechanism is responsible for preventing these galaxies from developing gas discs as symmetric as those of field galaxies.
\par

Taken together, these results indicate that the morphology of cold-gas discs at z=0 retains signatures of the environmental processes associated with filaments. Filament galaxies show systematically larger gas-disc asymmetries than field galaxies of the same stellar mass, consistent with the influence of the cosmic web tidal field even in systems that continue to accrete gas. At the same time, galaxies that experienced starvation do not simply stop accreting gas, but also retain the high level of asymmetry characteristic of earlier cosmic epochs. Finally, stripped galaxies display strongly enhanced asymmetries, as expected, further supporting the presence of cosmic web stripping.

\section{Gas temperature profile of filament galaxies}
\label{app:temp_profile}

\begin{figure}[h!]
    \centering
    \includegraphics[width=1\linewidth]{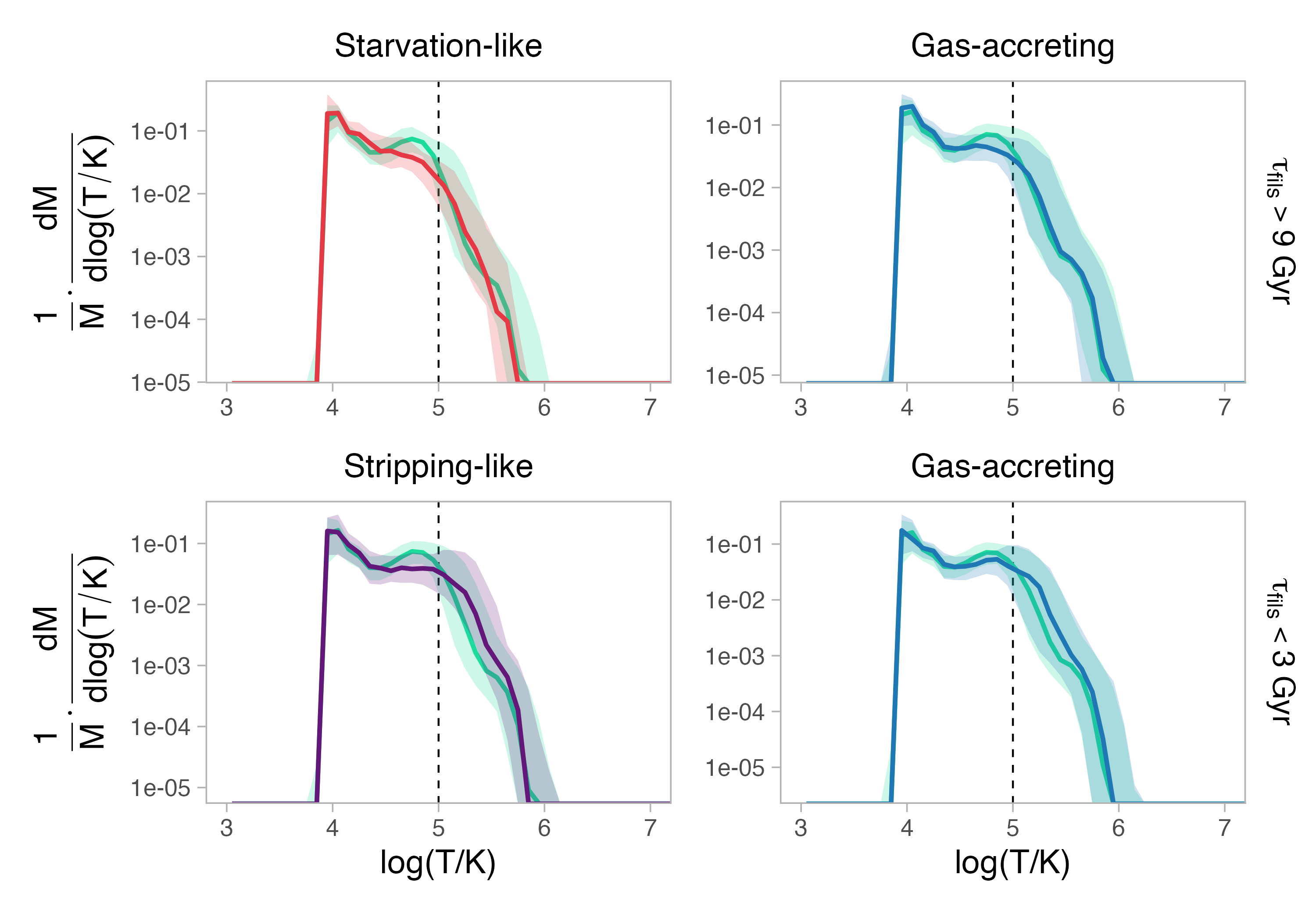}
    \caption{Normalised gas temperature distributions at z=0 for filament galaxies, split by gas-evolution scenario, compared to their stellar- and halo-mass-matched field counterparts. Curves show the median mass-weighted profiles, $(1/M)~\mathrm{d}M/\mathrm{d}\log(T/\mathrm{K})$, with shaded regions indicating the $16$th--$84$th percentiles. The vertical dashed line marks $T=10^5~\mathrm{K}$. }
    \label{fig:temp_profiles_appendix}
\end{figure}

To test whether the smaller sizes of cold gas discs in filament galaxies could arise because their gas is systematically hotter, we examine the gas temperature distribution at z=0 for galaxies that demonstrate different gas evolution scenarios in filaments and for their stellar- and halo-mass-matched field counterparts. For each galaxy, we compute the mass-weighted temperature profile, $(1/M)~\mathrm{d}M/\mathrm{d}\log(T/\mathrm{K})$, using all gas cells bound to the subhalo, and then derive the median profile and the $16$th--$84$th percentile range for each subsample. As shown in Fig.~\ref{fig:temp_profiles_appendix}, the temperature profiles of filament and field galaxies are very similar in all panels, indicating that filament galaxies do not exhibit a systematic excess of hot gas; therefore, the smaller $R_{90\%,\mathrm{cold~gas}}$ in filaments cannot be explained simply by an overall heating of the gas. Also, the fraction of cold gas ($T < 10^5~\mathrm{K}$) is very similar between filament and field galaxies in all subsamples. For $\tau_{\mathrm{fils}}>9~\mathrm{Gyr}$ starvation-like galaxies, we find $97.4^{+1.8}_{-4.7}\%$ compared to $97.5^{+1.8}_{-13.7}\%$ in the field, and for gas-accreting galaxies $94.6^{+4.3}_{-16.5}\%$ compared to $94.7^{+4.1}_{-23.2}\%$. For $\tau_{\mathrm{fils}}<3~\mathrm{Gyr}$, stripping-like galaxies have $91.7^{+5.1}_{-24.7}\%$ compared to $94.4^{+4.3}_{-20.3}\%$ in the field, while gas-accreting galaxies show $91.3^{+7.2}_{-21.0}\%$ compared to $94.3^{+4.4}_{-23.2}\%$. Overall, the cold gas fractions are consistent between filament and field galaxies within the uncertainties in all cases.

\end{appendix}
\end{document}